\documentclass[runningheads]{llncs}

\usepackage{eccv}
\usepackage{lmodern}
\usepackage{eccvabbrv}

\usepackage{graphicx}
\usepackage{booktabs}
\usepackage{gensymb}
\usepackage{adjustbox}
\usepackage{tabularx}
\usepackage{multirow}
\usepackage{placeins}
\usepackage{hyperref}

\newcommand\samethanks[1][\value{footnote}]{\footnotemark[#1]}

\begin{document}

\title{Helios 2.0: A Robust, Ultra-Low Power Gesture Recognition System for Event-Sensor based Wearables}
\titlerunning{Helios 2.0: Ultra-Low Power Gesture Recognition}

\author{Prarthana Bhattacharyya\thanks{These authors contributed equally.} \and
Joshua Mitton\samethanks \and Ryan Page\samethanks \and Owen Morgan\samethanks \and
Oliver Powell\samethanks \and Benjamin Menzies \and Gabriel Homewood \and
Kemi Jacobs \and Paolo Baesso \and Taru Muhonen \and Richard Vigars \and
Louis Berridge}
\authorrunning{P.~Bhattacharyya, J.~Mitton, R.~Page, O.~Morgan, O.~Powell et al.}
\institute{Ultraleap Ltd., Bristol, UK}

\maketitle
\vspace{-0.7cm}

\begin{abstract}
We present an advance in machine learning powered wearable technology: a
mobile-optimised, real-time, ultra-low-power gesture recognition model.
This model utilises an event camera system that enables natural hand gesture
control for smart glasses.
Critical challenges in hand gesture recognition include creating systems that
are intuitive, adaptable to diverse users and environments, and
energy-efficient allowing practical wearable applications.
Our approach addresses these challenges through four key contributions: a novel
machine learning model designed for ultra-low-power on device gesture
recognition, a novel training methodology to improve the gesture recognition
capability of the model, a novel simulator to generate synthetic micro-gesture
data, and purpose-built real-world evaluation datasets.
We first carefully selected microgestures: lateral thumb swipes across the
index finger (in both directions) and a double pinch between thumb and index
fingertips.
These human-centred interactions leverage natural hand movements, ensuring
intuitive usability without requiring users to learn complex command sequences.
To overcome variability in users and environments, we developed a simulation
methodology that enables comprehensive domain sampling without extensive
real-world data collection.
Our simulator synthesises longer, multi-gesture sequences using Markov-based
transitions, class-balanced sampling, and kinematic blending.
We propose a sequence-based training approach to learn robust micro-gesture
recognition entirely from simulated data.
For energy efficiency, we introduce a five-stage, quantisation-aware
architecture with >99.8\% of compute optimised for low-power DSP execution.
We demonstrate on real-world data that our proposed model is able to generalise
to challenging new users and environmental domains, achieving F1 scores above
80\%.
The model operates at just 6-8 mW when exploiting the Qualcomm Snapdragon
Hexagon DSP.
In addition, this model surpasses an F1 score of 80\% in all gesture classes in
user studies.
This improves on the state-of-the-art for F1 accuracy by 20\% with a power
reduction 25x when using DSP.
This advancement for the first time brings deploying ultra-low-power vision
systems in wearable devices closer and enables natural, always-on hand-based
interaction on smart glasses.
A real-time video demonstration of Helios 2.0 can be found \href{https://0e84f9dd10852326-tracking-platform-shared-public-assets.s3.eu-west-1.amazonaws.com/IMG_6222.mov}{at this link}.

\end{abstract}

\section{Introduction} \label{sec:intro} Hand gestures represent one of the most
natural human-computer interaction paradigms, offering intuitive control without
requiring visual attention from the user. As the number of smart glass devices
has exploded over the last year, highlighted by their emergence as a wearable
product category at the
\href{https://www.forbes.com/sites/timbajarin/2025/01/09/ces-2025-ai-health-wearables-and-smart-glasses-take-center-stage/}{Consumer
Electronics Show (CES) 2025}, efficient gesture interaction has become
increasingly important. Currently, these devices almost exclusively use
capacitive touch on the glasses' temple for their human-machine interface. This
solution however is often awkward to use, limits interaction vocabulary, and can
be socially conspicuous as users repeatedly reach to touch their eyewear.  The
industry has acknowledged these limitations by introducing additional
peripherals, such as rings or wrist bands. While these accessories enable more
intuitive and ergonomic hand-based input, they require users to wear and
maintain an additional piece of hardware. To overcome this need for additional
devices while still leveraging natural hand-based input, an event-based vision
system was previously proposed by Helios 1.0
\cite{bhattacharyya2024heliosextremelylowpower}.
\begin{figure}[t]
  \centering
  \begin{subfigure}[b]{0.48\linewidth}
    \centering
    \includegraphics[height=4.2cm]{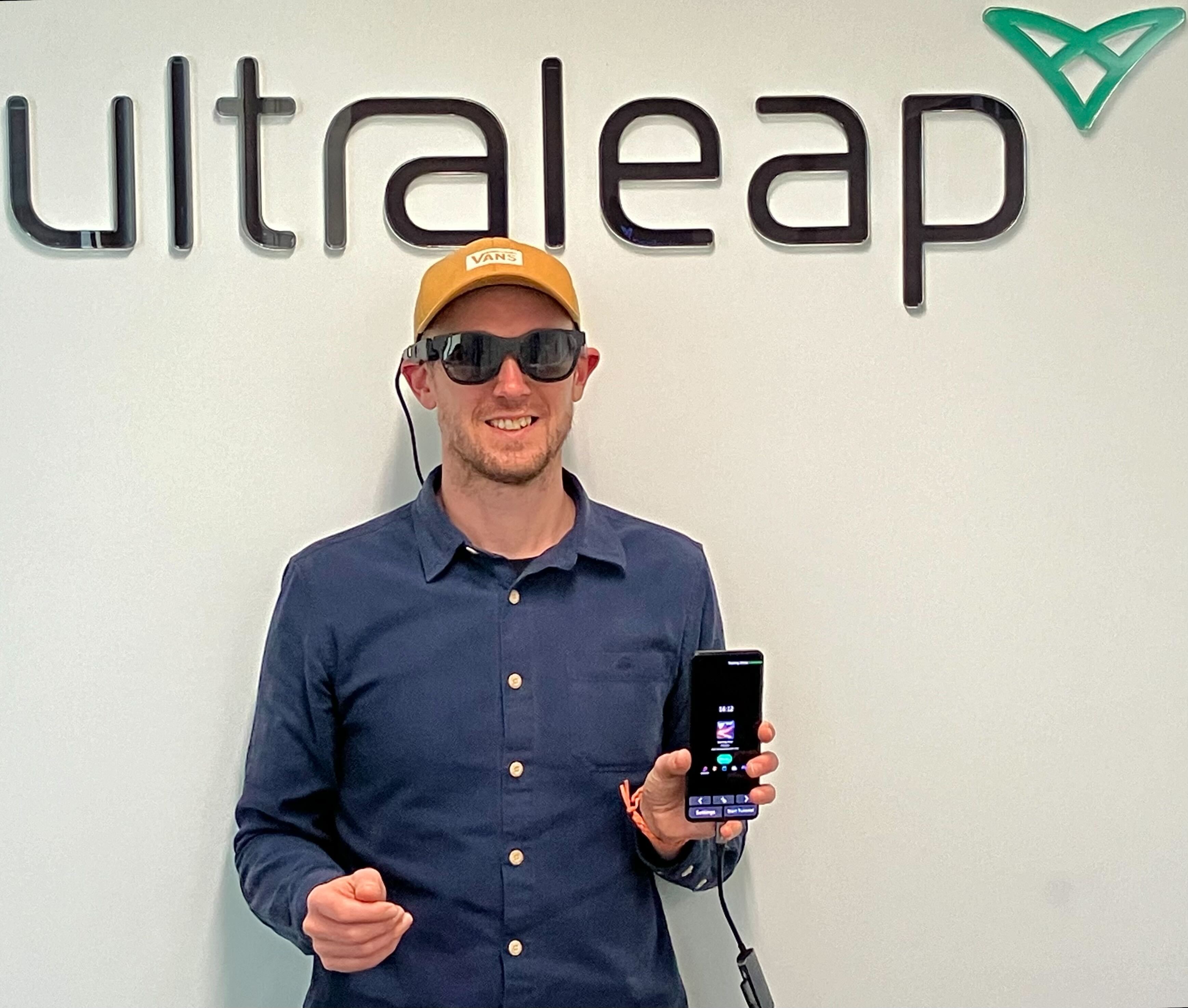}
    \caption{Real-time testing application.}
    \label{fig:teaser_fig}
  \end{subfigure}
  \hfill
  \begin{subfigure}[b]{0.48\linewidth}
    \centering
    \includegraphics[height=4.2cm]{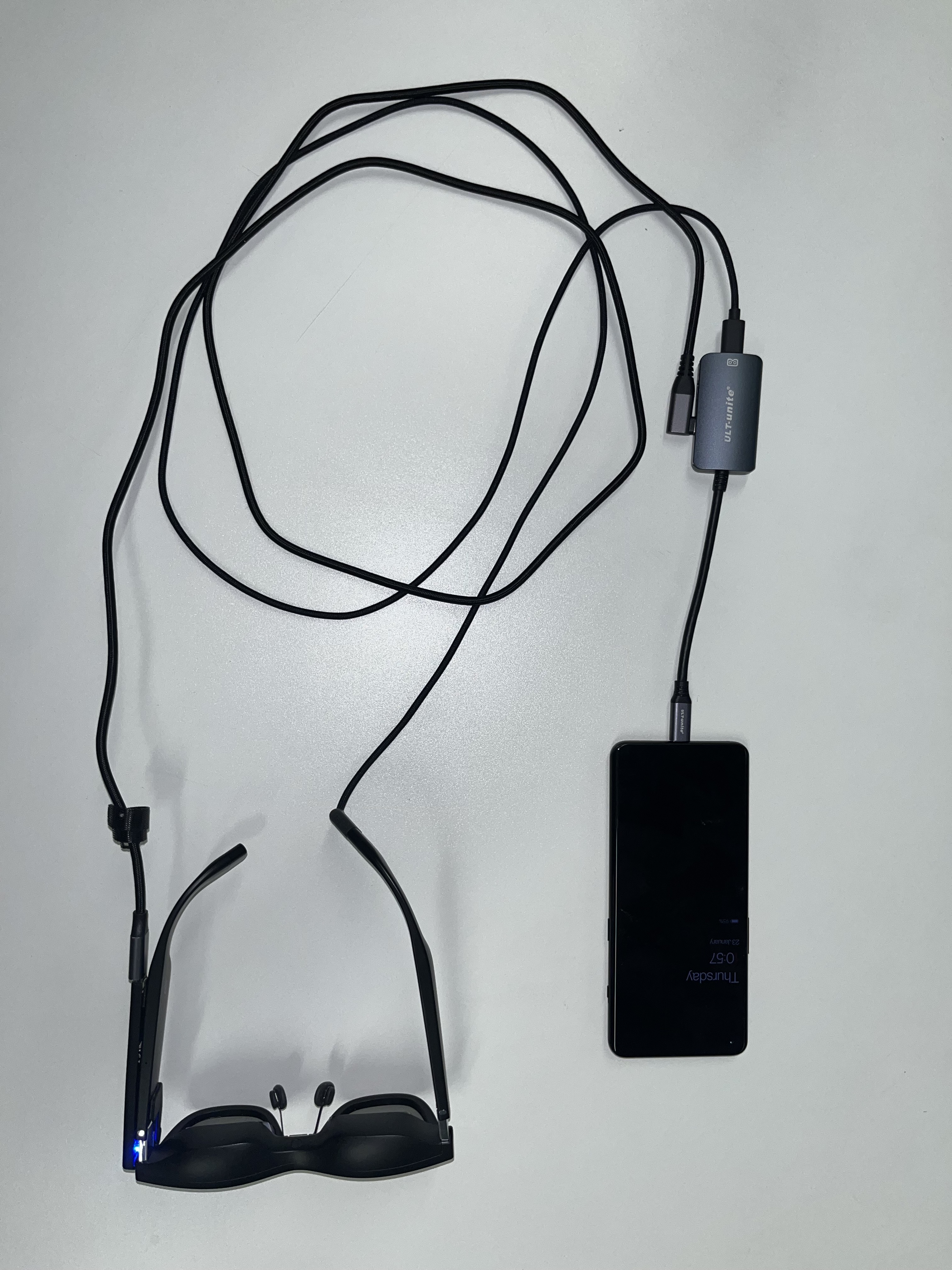}
    \caption{Event-camera glasses hardware.}
    \label{fig:xrealglasses}
  \end{subfigure}
  \caption{Helios 2.0 hardware and real-time testing setup. (a) A user running
  the machine learning model in real time. (b) Hardware used for data collection
  and model testing: the event camera is mounted on the left of the glasses,
  with the display connection on the right.}
  \label{fig:teaser_combined}
\end{figure}
In this work, we present significant improvements to the event-based gesture
recognition approach proposed in Helios 1.0. We develop a model that is
specifically designed to be ultra-low-power to enable always on smart glasses.
Through this we pay attention to the device each section of the model intends to
run on, ensuring that the majority of the models compute is quantised and run on
a DSP device specifically optimised for this.  In addition, we developed a novel
simulation method that makes it possible to create large diverse training
datasets without the need for expensive time consuming real-world data
collection. We also develop custom training pipelines that allow us, in addition
to our improved training datasets, to develop state-of-the-art gesture
recognition models. This is achieved through a new method to train with
multi-gesture temporal sequences, quantisation-aware training (QAT) and
fine-tuning on tailored synthetic datasets. These improvements result in a model
that achieves F1 accuracy greater than 70\% for the single-window (SW) model and
greater than 80\% for the temporal (T) model across multiple users and
environments, while
only consuming 6-8~mW when running computationally intensive portions on a
Qualcomm Snapdragon Hexagon DSP.  The model, which is specifically designed for
low-power operation, is composed of five stages, with two stages containing the
majority of the computational burden running on the DSP. With a model latency of
2.35~ms, the system enables smooth and responsive interaction that feels natural
to users. \Cref{fig:teaser_fig} depicts a user with Helios 2.0 hardware where
the proposed machine learning model runs in real time.

These advances make hand gesture interaction with smart glasses more accessible,
eliminating the need for additional accessories while maintaining a socially
acceptable and power-efficient interaction paradigm. The rest of the paper
breaks down the specific enhancements made to the system and presents
comprehensive evaluation results, before concluding with future directions for
this technology.

Our key contributions are as follows: \begin{enumerate}
    \itemsep0em
    \item \textbf{An ultra-low-power gesture-recognition model.} A five-stage,
    quantisation-aware architecture whose two heaviest stages run on a DSP,
    cutting power by 25$\times$ and improving F1 by 20\% over prior work;
    together these make always-on operation on smart glasses feasible.
    \item \textbf{A sequence-based training methodology.} Multi-gesture temporal
    sequences, quantisation-aware training (QAT) and rotational fine-tuning,
    which together let the model learn robust microgestures and resist false
    positives caused by head and hand ego-motion.
    \item \textbf{An event-camera microgesture simulator.} A simulation
    methodology that generates large, diverse training data without expensive
    real-world collection, allowing the model to be trained entirely in
    simulation.
    \item \textbf{Real-world evaluation benchmarks.} Purpose-built datasets
    spanning user, scene and outdoor variability, giving a more rigorous test of
    generalisation than prior work.
\end{enumerate}

Our work builds on hand gesture recognition, event-based vision, event-camera
datasets and simulators, and low-bit quantisation. The closest event-based
gesture systems either target large, deliberate gestures with the bulky DVS-128
camera \cite{GesturewithEvents_2017} or, like Meta's frame-based approach
\cite{STMG}, rely on explicit hand-skeleton tracking; both are ill-suited to
always-on smart eyewear. General event-based pipelines using dense CNNs or graph
networks remain too slow for on-device use (e.g.\ 202\,ms per inference on a
desktop GPU \cite{SparseEventGraphGNN}). Helios 2.0 recognises subtle
microgestures directly from raw events at 2.35\,ms and 6--8\,mW. A full
discussion of related work is provided in the supplementary material.

\section{Hardware} \label{sec:hardware}
This section details our test platform and power measurement setup.  
\subsection{Test Platform} Our system uses a Prophesee GenX320 event camera with
a custom USB readout PCB, based on Helios 1.0
\cite{bhattacharyya2024heliosextremelylowpower}. The key differences are
where the camera was mounted and execution hardware.  We mounted the event
camera on XReal Airs to enable richer visual experiences and user testing.
\Cref{fig:xrealglasses} shows this configuration, with the event camera and USB
readout on the temple, alongside the XReal's display connection. Both connect to
a USB hub linked to an Android phone.  For mobile-optimised model development,
we used an Android OnePlus 10 Pro with the Qualcomm Snapdragon 8 Gen1, featuring
Hexagon DSP for acceleration. While this offers more computing power than in
Helios 1.0, it provided insights into how DSP and NPUs would enhance model
performance in future smart glass products.
\subsection{Model Power Evaluation}
To determine the model power consumption a Snapdragon XR2Gen2 based platform was
used. The CPU architecture is similar to the Snapdragon 8 Gen1 featuring Kryo
Prime, Gold and Silver ARM based CPU cores and Hexagon
DSP~\cite{qualcommSoCpage}. The power was calculated by measuring the current
consumption using a 16-bit ADC that sampled the voltage across a shunt resistor
in series with the device. This was multiplied by the input voltage provided by
a dedicated PSU, also sampled by a 16-bit ADC, to give the total device power.
To calculate the power used by the model, the following test sequence is carried
out using a test harness: first the harness loads the input data for testing and
runs all base tasks, except for model execution, recording the power $P_{base}$.
On the next run, the harness runs and loads the same data, but this time
completes $N$ iterations of the model over it. The test harness then terminates
the measurement, logging the total power of the run, $P_{total}$. It should be
noted the data is from a user recording, so it is representative of what the
model would process at inference time. Finally, the power consumption of the
model $P_{model}$ was calculated using the following $P_{model}=(P_{total} -
P_{base})/N$.

\section{Datasets} \label{sec:dataset} \subsection{Synthetic Datasets} This
section explains our simulation-based approach to dataset generation. By
combining eSIM's event generation capabilities \cite{Rebecq18corl} with a
custom Unity-powered \cite{unity} rendering engine, our system synthesises
realistic hand gesture events for training purposes.
\subsubsection{Simulator Details} 
\begin{figure}[t]
    \centering
    \begin{subfigure}[t]{0.48\linewidth}
	\centering
	\includegraphics[width=0.8\linewidth]{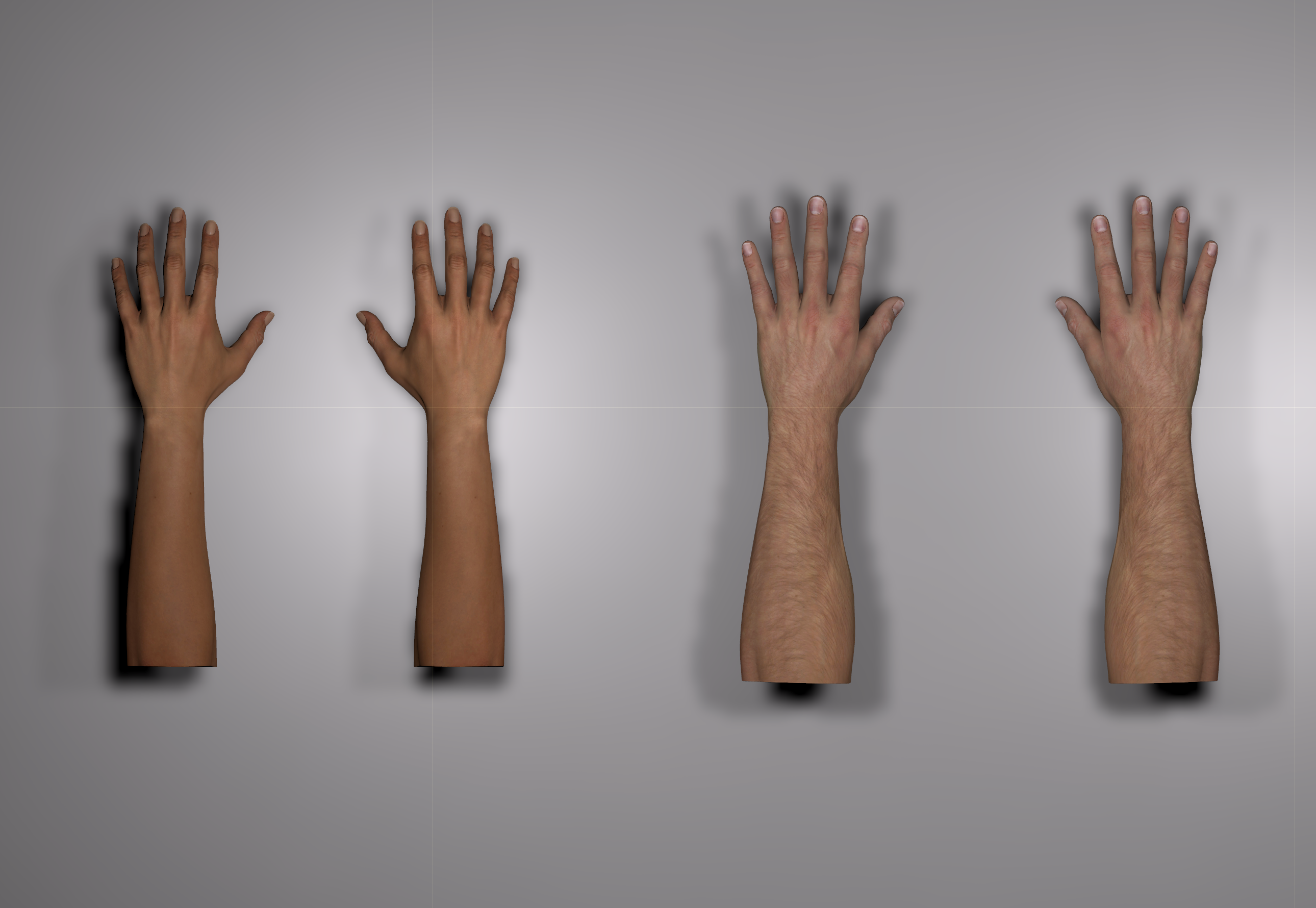}
	\caption{3D hand models used for simulation.}
	\label{fig:saltandpepper}
    \end{subfigure}
    \hfill
    \begin{subfigure}[t]{0.48\linewidth}
	\centering
	\raisebox{0.2cm}{\includegraphics[width=0.9\linewidth]{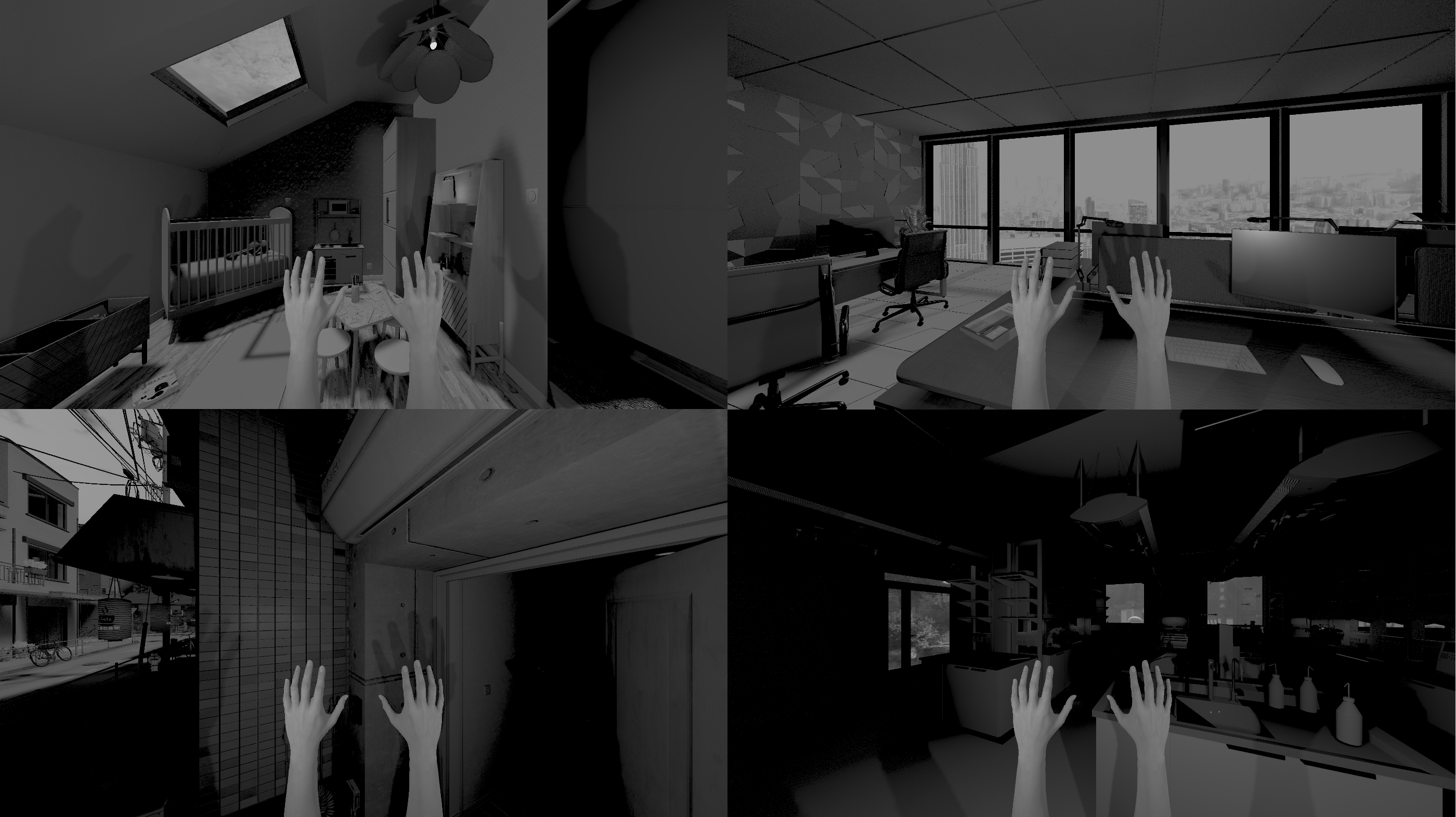}}
	\caption{3D environments used in simulation.}
	\label{fig:simbackgrounds}
    \end{subfigure}
    \caption{Simulation assets used in Helios 2.0: 3D hand models and diverse
    3D environments.}
    \label{fig:sim_assets}
\end{figure}

The simulator uses data from a commercial
\textit{Orion} \cite{ul-orion} hand-tracking system, recorded at 90\,Hz and
segmented into 40-frame sequences. Pre-processing balances left/right chiralities
by mirroring, removes poor or repeatedly untracked poses, and fills single
missing frames by linear interpolation to avoid rendering artifacts. The
processed data drives hand-mesh rigging (\Cref{fig:saltandpepper}) in Unity's
High Definition Render Pipeline, producing photorealistic hands in synthetic 3D
environments (\Cref{fig:simbackgrounds}); \Cref{fig:datasetbalance} shows the
class distribution of the resulting training set.
\subsubsection{Simulator Augmentation} Our augmentations fall into three
categories: environment modifications (lighting and hand positioning), random
mutation of the recorded hand-tracking data, and formation of the data into
Helios 2.0 microgesture classes. We vary scene lighting (50-400\% of nominal) at
fixed camera exposure, and move the camera along a path-finding route between
`safe' zones on the $(x, y)$ ground plane while rotating about its yaw axis to
mimic the head motion of head-mounted AR glasses. For per-frame diversity, we
perturb a forward-kinematic bone-angle model with small random deviations, so
each run yields unique data.
\subsubsection{Simulation of Longer Multiple Micro Gesture Sequences}
\label{ssec: longer-sequence-datasets} 
\begin{figure}[t]
    \centering
    \begin{subfigure}[t]{0.48\linewidth}
	\centering
	\includegraphics[width=0.7\linewidth]{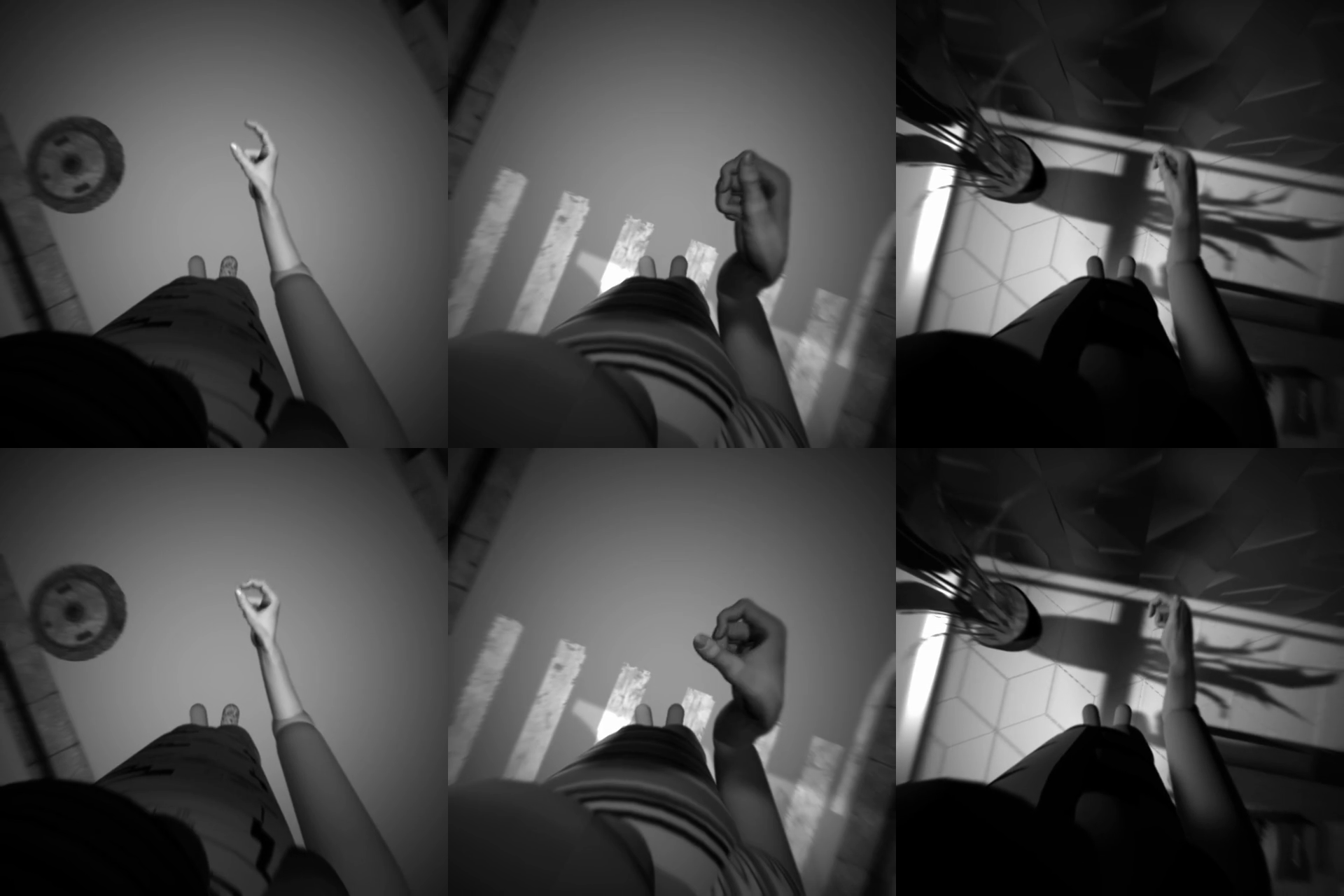}
	\caption{Left to right: pinch, left swipe, right swipe. The end point of
	the microgesture is illustrated at the bottom.}
	\label{fig:uxgestures}
    \end{subfigure}
    \hfill
    \begin{subfigure}[t]{0.48\linewidth}
	\centering
	\raisebox{0.5cm}{\includegraphics[width=0.8\linewidth]{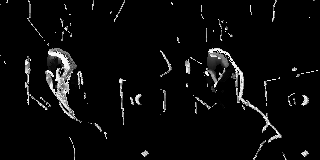}}
	\caption{Example time surface with left hand side illustrating the
	positive polarity image and right hand side illustrating the negative
	polarity image.}
	\label{fig:example-timesurface}
    \end{subfigure}
    \caption{Gestures modelled for smart interactions by Helios 2.0 and an
    example of time surface representation.}
    \label{fig:gesturesmodeledandtimesurface}
\end{figure}

In Helios 1.0
\cite{bhattacharyya2024heliosextremelylowpower}, a 7-class model was proposed
with gestures comprising (1) unknown motions, (2) untracked hands, (3) pinch,
(4) double pinch, (5) swipe left, (6) swipe right, and (7) rest positions,
generated in a single time window of 434~ms. In Helios 2.0, to improve system
robustness, we extended sequence length from 434ms to 2s and expanded from 7 to
10 classes by adding "return" versions of several gestures. The expanded
classification includes: (1) Unknown, (2) Untracked, (3) Pinch, (4) Double
Pinch, (5) Pinch (return), (6) Swipe Left, (7) Swipe Left (return), (8) Swipe
Right, (9) Swipe Right (return), and (10) Rest. \Cref{fig:uxgestures}
illustrates swipe and pinch gestures in the simulator.
Our animation uses forward kinematics with motion derived from Optitrack user
studies, which revealed gesture motions approximating sigmoid curves $S(x) =
\dfrac{1}{1 + e^{-mx}}$; we implement four variations of the steepness $m$ for
diversity. Together with the expanded gesture set, this addresses false
positives such as the return motion after a right swipe being misread as a left
swipe. We cap each gesture at ${\sim}333$\,ms, allowing up to six gestures within
a 2\,s sequence. Rather than random selection, we implemented structured transition rules between
gestures as a Markov chain, with rest position serving as a root state. For
class balance, we manually weighted probabilities (\Cref{fig:datasetbalance})
rather than using algorithmic approaches like Reverse Page Rank
\cite{Berend2020}. Between certain transitions (rest to swipe, rest to unknown,
or rest to pinch), we apply frame blending through linear interpolation in angle
space to ensure kinematically plausible motions.

\subsubsection{Rendering and Event Generation} \label{finalsimoutput} After
synthesising sequences of hand positions, we first render frames and then use
them to generate events utilising an Unity implementation of eSIM
\cite{Rebecq18corl}. An event is given by $P_{i}:=(x_{i}, y_{i}, p_{i},
t_{i})$. Here, $x_{i}, y_{i}$ are the $(x, y)$ pixel coordinates respectively,
$p_{i}$ is the polarity defined as 1 for positive changes and 0 for negative
changes and $t_{i}$ is the time the event occurred, in $ns$ for event $i$. To
ensure contrast detection is possible across the scene, we employ High Dynamic
Range (HDR) rendering to prevent event loss in regions that saturate in the
8-bit range (0-255). This accommodates more diverse environmental and lighting
variations, closer to real-world scenarios. The simulator outputs events,
current active gesture labels and 3D hand joint location for precise event timing.  
\par To validate our
simulator, a dynamic target was created both in the lab and in our simulator.
The sim-to-real gap in event rate was first tuned on a high-contrast target. It
was then fine-tuned with real-world hand gesture data to the final contrast
ratio. 
being used to train models.
\subsection{Real Dataset Curation} We developed an Android testing application
to collect real world gesture data for model benchmarking. The app captured user
experience level, personal characteristics (hand size, skin tone using the Monk
Skin Tone Scale), and environmental factors (lighting, motion status).  The
tests consisted of 60 trials where users performed random prompted gestures
(pinch, left/right swipe) without visual feedback to prevent behaviour
modification. Each trial logged the requested gesture, timing information, and
raw event stream data for subsequent analysis.

Datasets from internal users are annotated with the Monk Skin Tone
Scale~\cite{mst} (illustrated in the supplementary material), giving a
qualitative measure of demographic coverage as the benchmarks grow.

Our data-collection procedure followed informed-consent and participant-privacy
practices; a full ethics statement is provided in the supplementary material.

\subsubsection{Human Variability Study} \label{ssec:h_var} 
\begin{figure}[t]
    \centering
    \includegraphics[height=2.5cm]{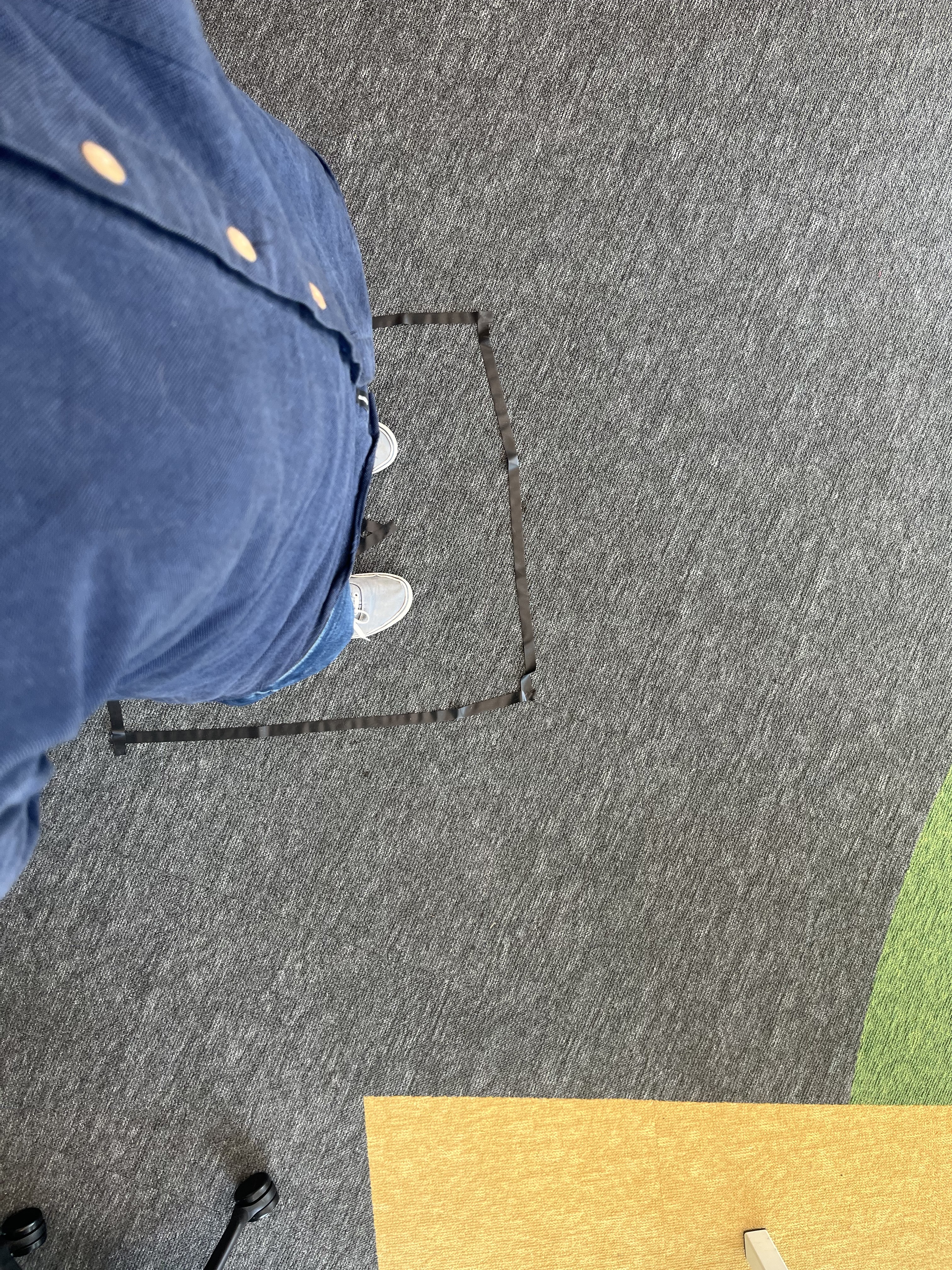}
    \caption{Background used for the Human Variability Study, with floor marking
    to ensure minimal background variability between users. Lighting was measured
    at 60~lux.}
    \label{fig:user_area}
\end{figure}

\begin{figure}[t]
    \centering
    \begin{subfigure}[t]{0.45\linewidth}
	\centering
	\includegraphics[width=\linewidth]{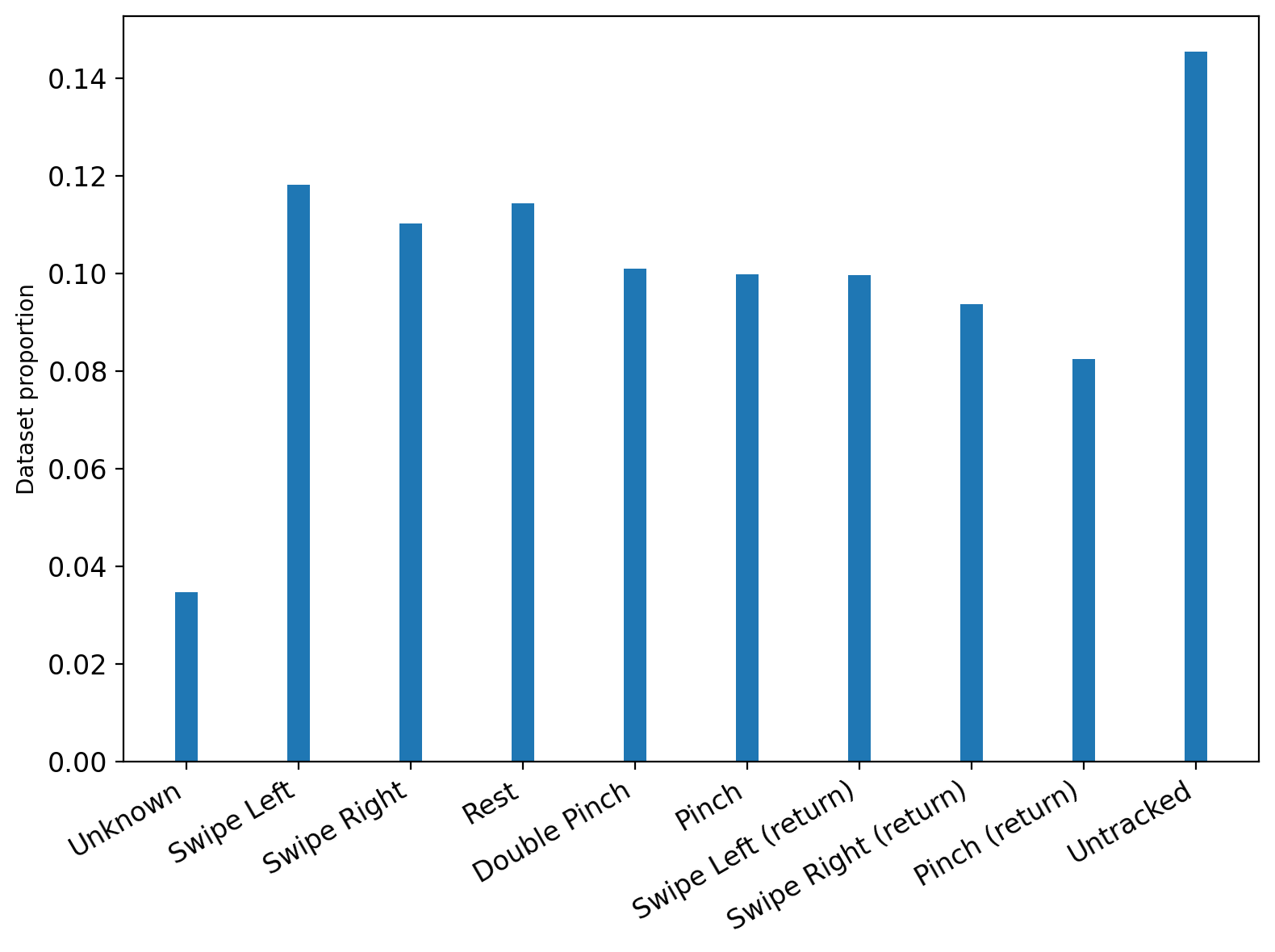}
	\caption{Distribution of classes from a subset of the simulated training
	data.}
	\label{fig:datasetbalance}
    \end{subfigure}
    \hfill
    \begin{subfigure}[t]{0.5\linewidth}
	\centering
	\raisebox{0.6cm}{\includegraphics[width=\linewidth]{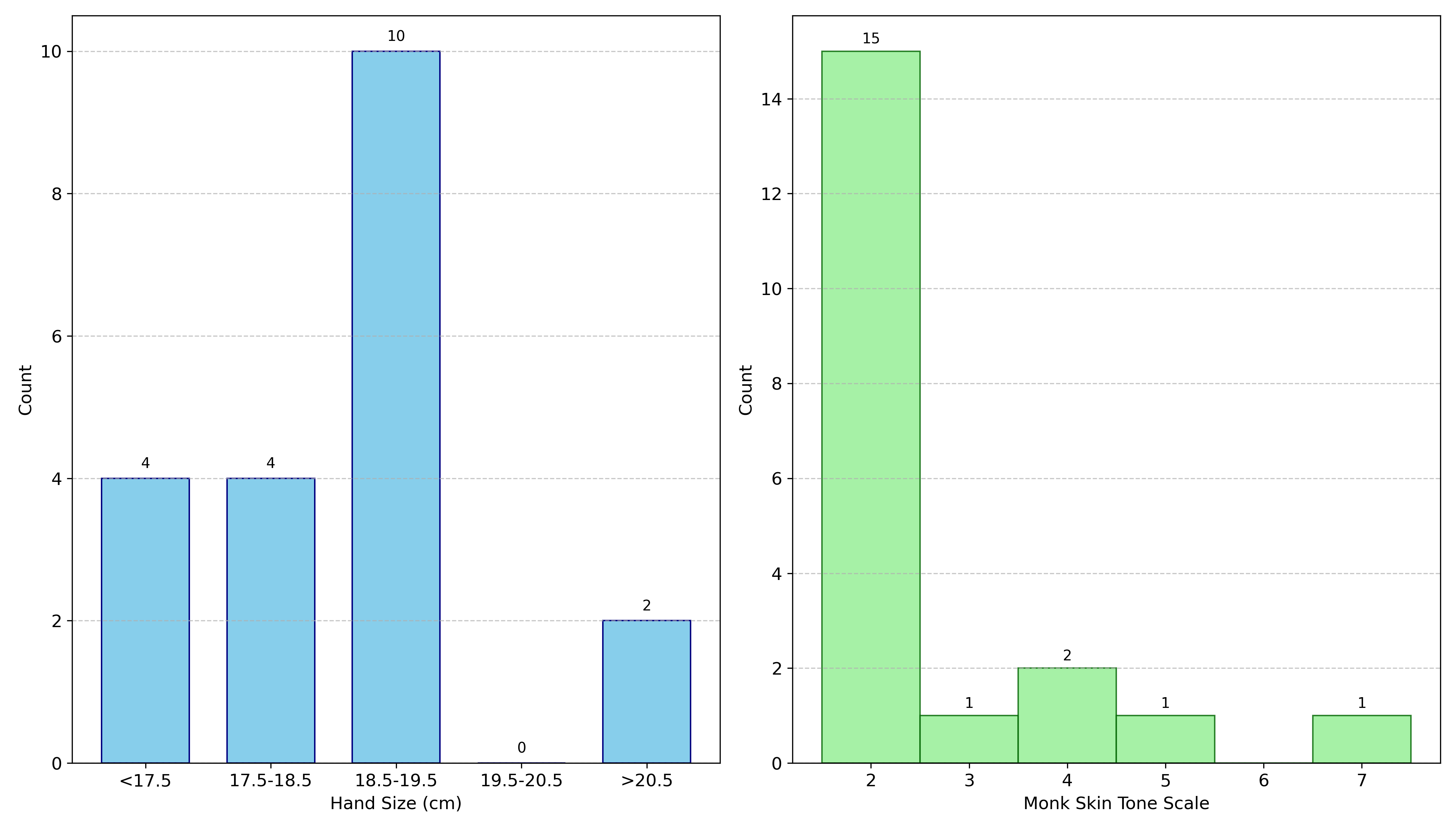}}
	\caption{Histograms of user details. Left shows distribution of hand size
	and right shows distribution of skin tone based on the Monk Skin Tone
	Scale~\cite{mst}.}
	\label{fig:userdist}
    \end{subfigure}
    \caption{Distribution of gestures in the training data, and a histogram of
    the user hand sizes in the testing data for the human variability study.}
    \label{fig:histograms}
    \vspace{-0.4cm}
\end{figure}

In
the Human Variability Dataset, the user was asked to stand within a square and
were oriented using an arrow marked on the floor. For the study, an RGB image
from the view point of the event camera is shown in \Cref{fig:user_area}. This
ensured that between users the background did not change. Twenty users with
varying experience levels completed the testing protocol under controlled office
conditions. These twenty users were divided into four groups for systematic data
collection. In \Cref{fig:userdist}, we show the distribution of hand sizes
(measured from the base of the wrist to tip of the middle finger) and skin tones
among participants. Video recordings supplemented quantitative metrics to
qualitatively analyse low-performing cases.
\subsubsection{Scene Variability Study} \label{ssec:s_var} 
\begin{figure}
    \centering
    \begin{subfigure}{0.23\textwidth}
	\includegraphics[width=\textwidth]{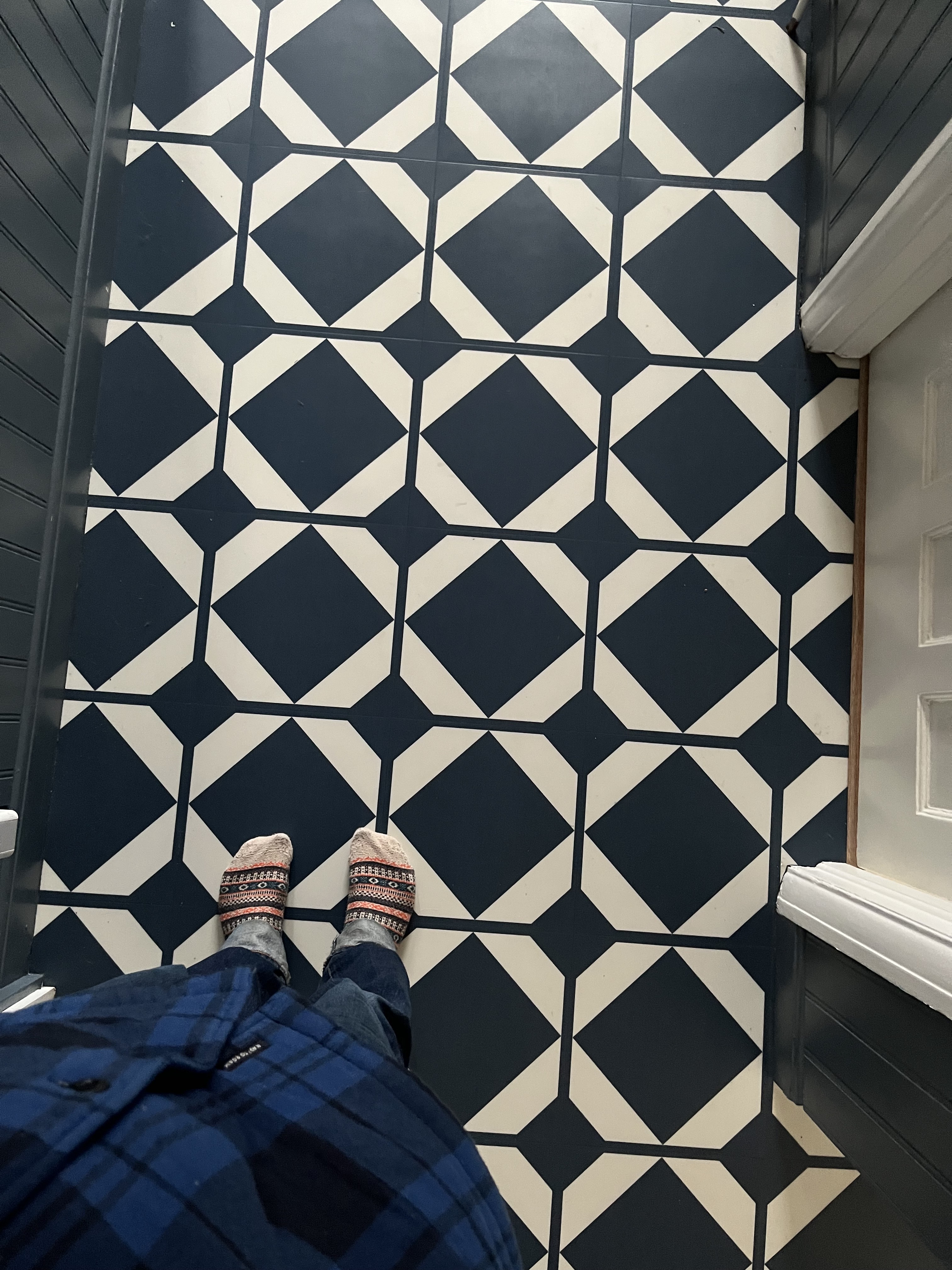}
	\caption{Hallway}
	\label{fig:hallway}
    \end{subfigure}
    ~ 
    \begin{subfigure}{0.23\textwidth}
	\includegraphics[width=\textwidth]{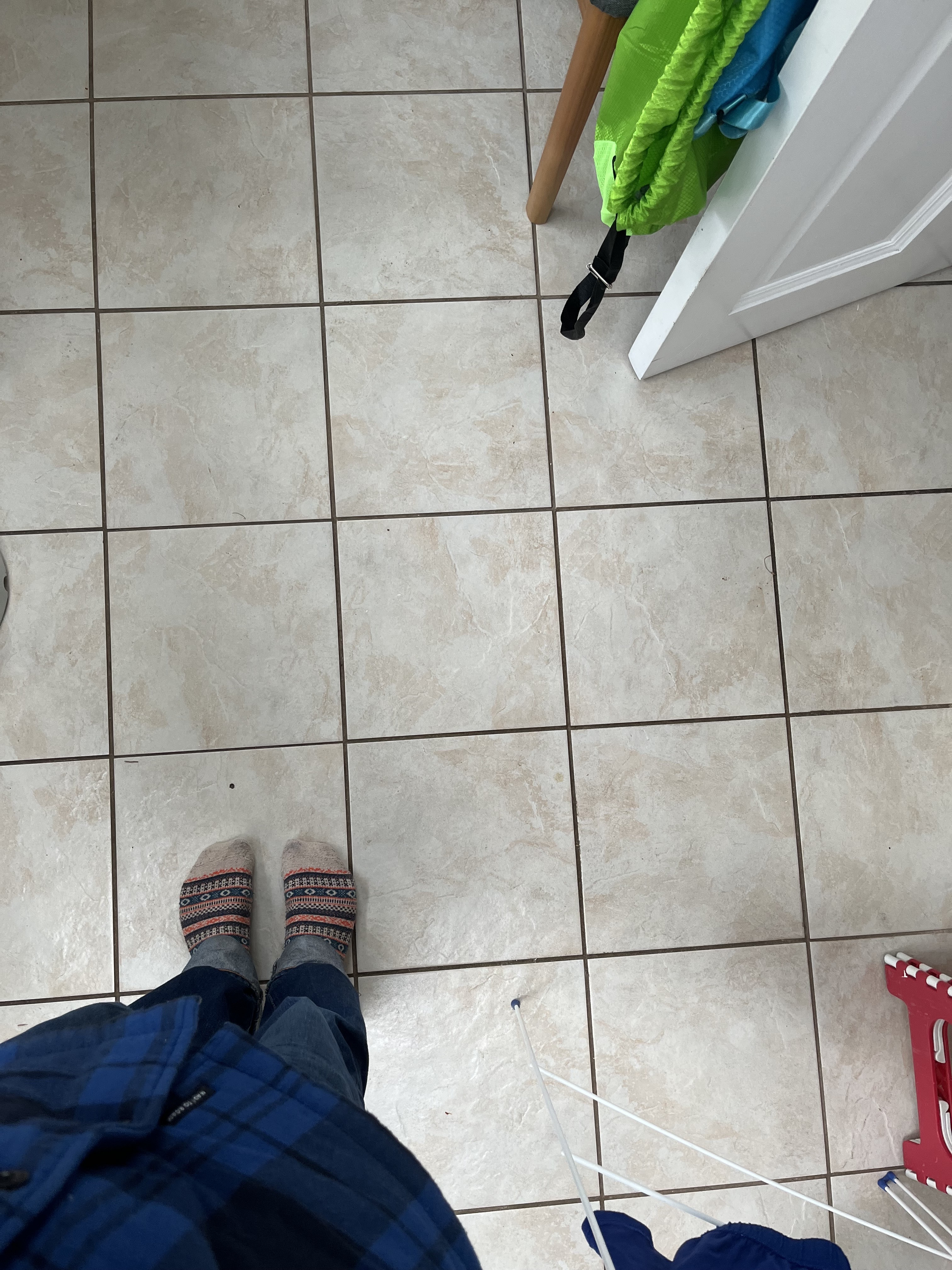}
	\caption{Utility}
	\label{fig:utility-room}
    \end{subfigure}
    ~ 
    \begin{subfigure}{0.23\textwidth}
	\includegraphics[width=\textwidth]{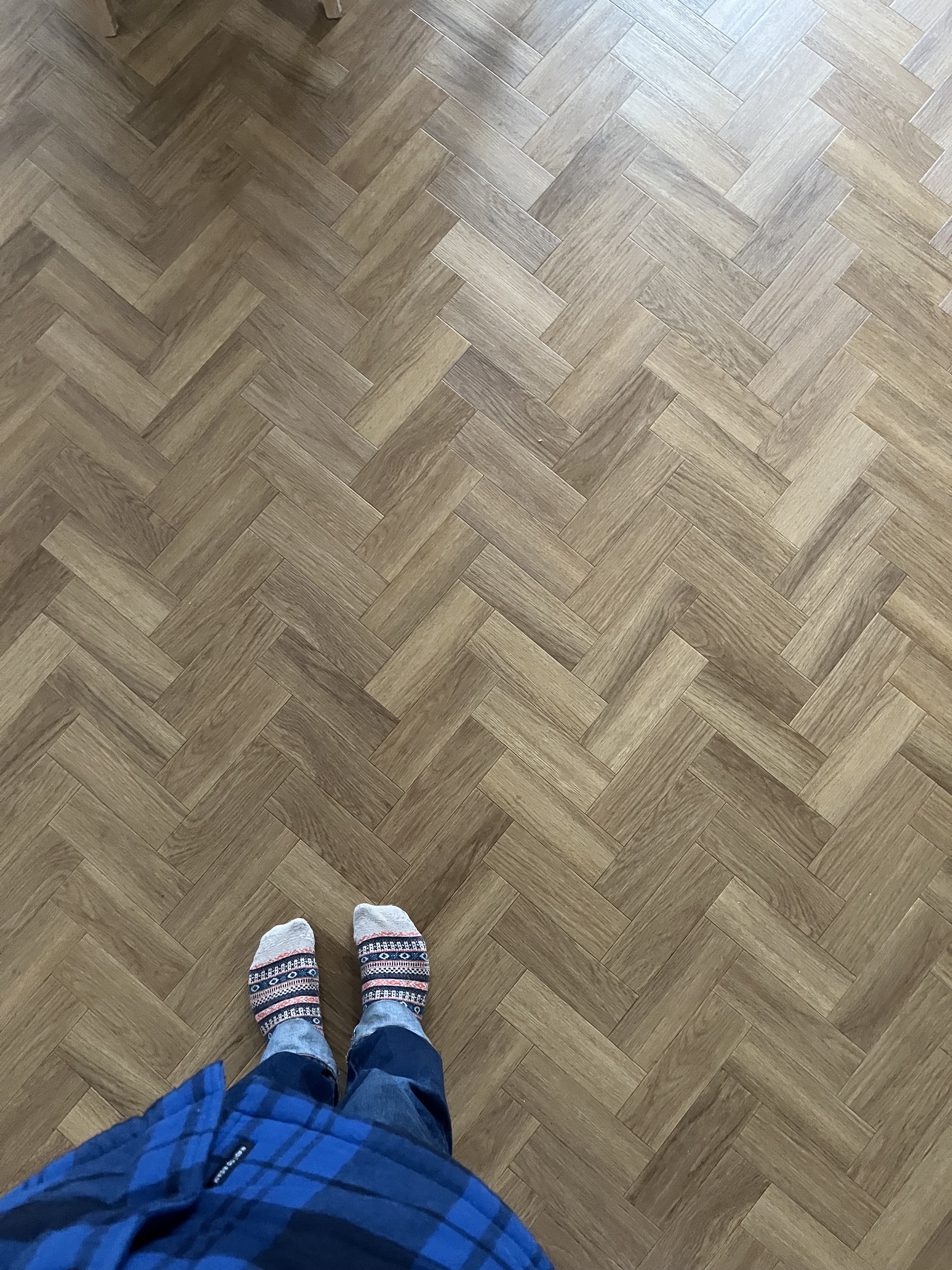}
	\caption{Kitchen 1}
	\label{fig:kitchen1}
    \end{subfigure}
    ~
    \begin{subfigure}{0.23\textwidth}
	\includegraphics[width=\textwidth]{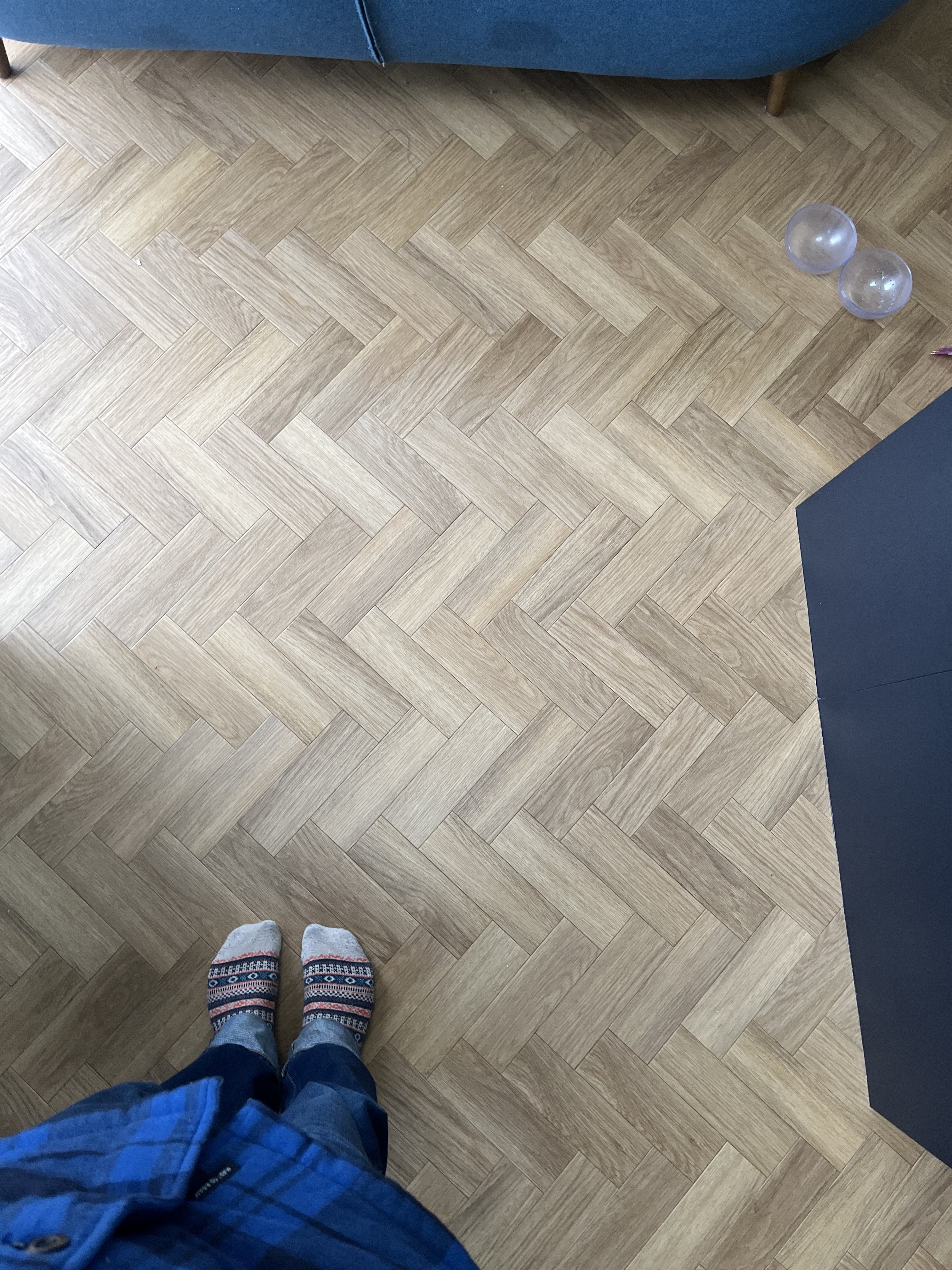}
	\caption{Kitchen 2}
	\label{fig:kitchen2}
    \end{subfigure}
    \caption{Various backgrounds with natural illumination measured at the event
    camera position: (a) 8~lux, (b) 140~lux, (c) 30~lux,
    (d) 240~lux.} \label{fig:backgrounds}
\end{figure}

To test across more
challenging environments, the Scene Variability Dataset was created. This
consists of different floor textures, with varying light levels.
\Cref{fig:backgrounds} illustrates the variation in backgrounds used. An expert
user conducted test sequences across four different home environments with only
natural lighting. These included various floor textures, with light levels
between 8-240~lux to test the model robustness.
\subsubsection{Outdoor Performance Study} \label{ssec:outdoor_var} 
\begin{figure}
    \centering \includegraphics[width=0.4\linewidth]{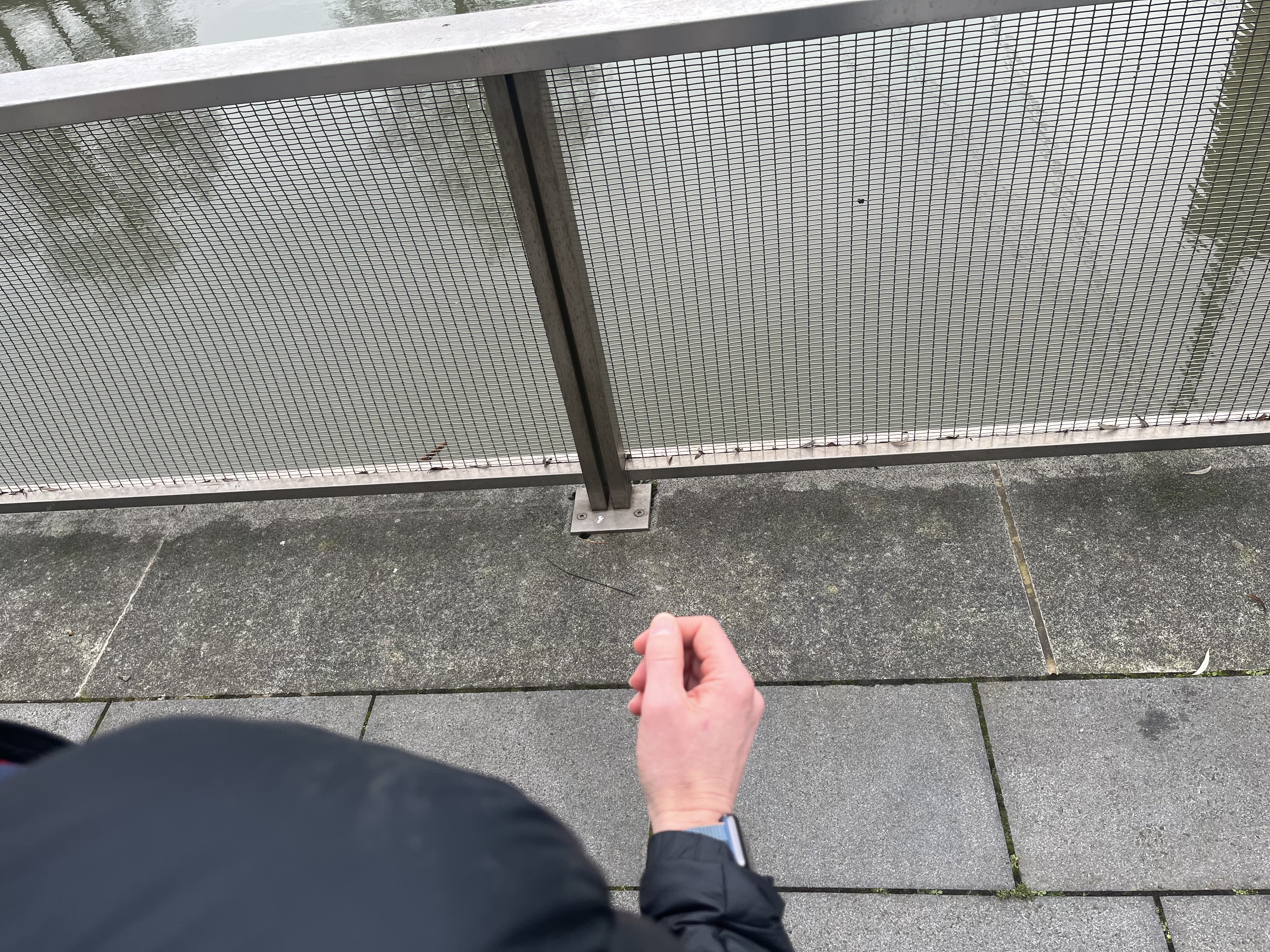}
    \caption{Outdoor scene showing hand in foreground with paving stones as a
    background. Ambient light= 3000~lux.} \label{fig:outdooors} \vspace{-0.3cm}
\end{figure}

In the final
dataset we consider an outdoor scene. \Cref{fig:outdooors} shows a RGB reference
image from this study. Group 2 participants from the human variability study
(from \cref{ssec:h_var}) repeated the same test sequence outdoors. Tests
occurred within a 30-minute window with consistent light levels of 3000 lux.

\section{Methodology} \label{sec:method} \subsection{Machine Learning Model}
\label{sec:ml}
We developed a mobile-friendly machine learning architecture for microgesture
detection optimised for low-power, low-latency performance on both CPU and DSP
targets. This makes it particularly well-suited to always-on smart eyewear.
\subsubsection{Event Representation for Helios 2.0} \label{sssec:
event-representation} In this section, we first describe how the event stream
generated in \cref{finalsimoutput} is processed before being used to train
models.  We propose to use polarity-separated event Time Surfaces (TS) as our
event representation for training. We build on the Locally-Normalised Event
Surfaces (LNES) approach from \cite{rudnev2021eventhands}. This achieves both
computational and storage efficiency. By separating polarities, we also mitigate
event clashes and ensure a more structured and accurate representation of the
temporal dynamics in event data. A key technical challenge is integrating the TS
representation with the synthetic dataset produced in \cref{finalsimoutput}.

\par To construct the TS representation \(\mathcal{I}_k \in \mathbb{R}^{w \times
h \times 2}\), we divide an event stream of length $L$ into discrete time windows of 240\,\text{ms}.
In our longer-sequence datasets, we set \(L = 2\,\text{s}\), with a step size of 80\,\text{ms} each,
giving rise to 25 steps in the sequence. So for $k=0$ events between
0-240\,\text{ms} are aggregated, then for $k=1$ events between 80-320\,\text{ms} are
aggregated and so on.
We denote the duration of each time
window by \(T_s\). $\mathcal{I}_k$ is created by first initialising it with
zeros and collecting events that have incoming timestamps $T_i$ within this
window by iterating through from the oldest to the newest event. We reduced
$T_s$ from 434ms to 240ms to improve temporal consistency for shorter gestures
in longer windows. To create LNES, we initialise the TS $\mathcal{I}_k$ to zeros
and update pixel values with an exponential decay: \begin{equation}
    \mathcal{I}_{k} = \left\{ \begin{array}{ll}
       e^{-\lambda\left({\dfrac{T_{max} - T_{i}}{T_s}}\right)}, & \mbox{if }
       T_{max} - T_{i} < T_{s} \\ 0,  &  \mbox{otherwise}
    \end{array} \right.  \label{eqn:time-surface}
\end{equation} Here $T_{max}$ is the upper time of the window,
$\lambda=5$ is the decay constant, and $T_i$ is the timestamp of the incoming
event. During inference, pixels with timestamps older than $T_s$ are reset to
zero. For an event stream with image bounds $\{w, h\}$, we accumulate positive
polarity events in the range $\{w, h\}$ and negative polarity events in the
range $\{w, 2h\}$, thereby reshaping \(\mathcal{I}_k \in \mathbb{R}^{w \times
2h}\) (\cref{fig:example-timesurface} - note for illustration purposes
Figure~\ref{fig:example-timesurface} displays the image as $\{2w, h\}$).  The 3D
hand joint locations are reprojected into 2D camera space to obtain bounding box
locations. Points beyond image bounds are clipped to $\{[0, w-1], [0, h-1]\}$.
Bounding box corners are computed via min-max operations, excluding points below
the wrist for better box tightness. As an improvement on prior work, we enforce
square bounding boxes for improved training accuracy and reduced sim-to-real gap
at inference time.
\subsubsection{Helios 2.0 Model Architecture and Quantisation} 
\begin{figure}[t]
  \centering \includegraphics[width=\textwidth]{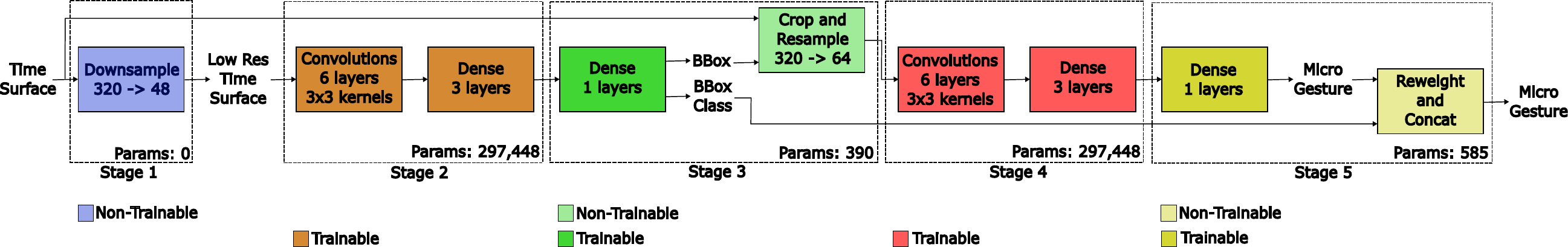}
  \caption{High-level block diagram of the five stage microgesture detection
  model architecture} \label{fig:model} \vspace{-0.3cm}
\end{figure}

\label{sssec:
model-architecture} Helios 2.0 introduces a five-stage, quantisation-aware
architecture with $>$99.8\% of compute optimised for low-power DSP execution
(\Cref{fig:model}). Each stage is designed for the following purposes: \begin{itemize}
    \item Stage 1: Downsamples the input time surface (TS) representation to a
    lower resolution. \item Stage 2: Comprises of convolutional
    layers and dense layers that perform feature extraction. The number of
    layers is specifically chosen to most efficiently extract the features by
    initially lowering the spatial resolution before predicting dense features.
    This layer contains a significant amount of the models parameters and is
    quantised to run efficiently on a DSP.  \item Stage 3: Has a non-quantised
    dense layer to predict floating point values for a bounding box, which
    ensures the model can predict the bounding box with high accuracy. Then
    stage 3 crops and resizes the original input using the predicted bounding
    box.  \item Stage 4: Comprises of convolutional layers and dense layers that
    perform feature extraction on the crop of the hand within the original
    image. This is again specifically chosen to most efficiently extract the
    features by initially lowering the spatial resolution before predicting
    dense features. This layer in addition to stage 2 makes up the majority of
    the models parameters and is quantised to run efficiently on a DSP.  \item
    Stage 5: Has a single dense layer to predict the micro gesture that is
    contained within the cropped image. It then combines stage 4 predictions
    with stage 2 hand-presence probabilities to produce final microgesture
    predictions.
\end{itemize} The CNN components in stages 2 and 4 account for $>$99.8\% of
parameters and floating point operations. We quantised these stages to run on a
Qualcomm Hexagon DSP, reducing power consumption and inference time. The final
dense layers of these stages and the remaining three stages (which include
interpolation and softmax functions) use 32-bit floating-point computation to
maintain accuracy, as their minimal compute requirements make quantisation
unnecessary. We implemented quantisation-aware training (QAT) with 8-bit
symmetric, per-channel weight quantisation and 8-bit asymmetric, per-tensor
activation quantisation. This scheme is natively supported by Hexagon DSP
hardware that balances accuracy, latency, and power consumption. QAT is
particularly important for our parameter-constrained model to mitigate
quantisation noise effects.

\subsubsection{Training Helios 2.0 with Proposed Simulated Dataset}
\label{sssec: training-with-longer-sequences} A key contribution of our work is
training the model proposed in \cref{sssec: model-architecture} with the longer
sequence dataset proposed in \cref{finalsimoutput} and their TS representations
proposed in \cref{sssec: event-representation}. We aim to learn the conditional
probability $p(g|\mathcal{I}_{k})$ of predicting gesture class $g$ given input
$\mathcal{I}_{k}$, where $\mathcal{I}_{k}$ is the TS representation for
time-step $k$ produced in \cref{sssec: event-representation}. Our datasets
feature six distinct gestures within a 2~s sequence, each lasting
$\sim$333~ms. As outlined in \cref{finalsimoutput} and \cref{sssec:
event-representation}, the simulator generates gesture labels. 
To obtain $g$, the labels must be aggregated across the 240\,\text{ms} window. 
For the first sample in a sequence, we select the maximum occurring label as $g$.
Subsequently, we assign a new label only if its
proportion exceeds the gesture threshold (e.g., at 0.6 threshold, a transition
from `rest' to `swipe right' requires at least 60\% of labels to be `swipe
right'). This results in a corresponding $g$ for each $\mathcal{I}_k$ generated. A model
trained on a single time surface $\mathcal{I}_k$ (two polarity channels,
dimensionality \{w, 2h\}) is our \emph{single-window} model (SW). To capture
broader temporal context, we also feed three consecutive TS samples
($\mathcal{I}_k$, $\mathcal{I}_{k-1}$, and $\mathcal{I}_{k-2}$) into the model,
effectively capturing gestural evolution (e.g., a `swipe left' should be
followed by a `swipe left return'). Including positive and negative polarities,
this representation has dimensionality \{w, 6h\}; we call this the
\emph{temporal} model (T), our primary, highest-accuracy configuration. The
single-window model trades a little accuracy for the lowest power and latency,
giving two operating points on one architecture. We provide more details about
our training and inference setup in the supplementary material.

\subsection{Finetuning Helios 2.0} \label{ssec: finetuning} To address the lack
of rotation invariance in CNN features, we augment our training dataset with
rotational variations and create an augmented finetuning dataset. While the
longer-sequence dataset (\cref{ssec: longer-sequence-datasets}) captures diverse
microgestures, it lacks comprehensive hand pose and orientation coverage. We
apply uniform random rotations (25\degree - 40\degree) to each 2~s sequence,
enhancing the model's robustness to rotational variance. During fine-tuning, we
freeze stages 1-3 and train only stages 4-5 with a reduced learning rate. This
preserves bounding-box localisation and low-level features while adapting to
rotational variations. We isolate rotation into this stage rather than augmenting
from the outset because rotating inputs during initial training destabilises the
stage-1--3 bounding-box regression; adapting only the stage-4--5 classifier adds
rotational robustness more cheaply and stably than retraining from scratch.

\section{Experiments} \label{sec:experimental_results}
Our headline result is that Helios~2.0 improves mean F1 by 20\% while reducing
power by 25$\times$ over Helios~1.0, operating at just 6--8\,mW on a Qualcomm
Snapdragon Hexagon DSP; the temporal (T) model is our most accurate
configuration and the single-window (SW) model our lowest-power one. We first
evaluate our
single-window (SW) low-power QAT model trained from scratch on simulated data
(90\% training, 10\% validation) in \cref{sec:performancesimulateddata}. Unless
otherwise stated, all models are trained on our largest longer-sequence dataset
(\cref{finalsimoutput}), denoted $4\times$, with $\mathord{\sim}100k$ training
samples per microgesture class. We also apply the rotation augmentation described
in \cref{ssec: finetuning} to create its augmented variant, $4\times$ Augmented.

\par In \cref{sec:2channel_6channel_cpu_dsp}, we then assess both the
single-window (SW) and temporal (T) models across both CPU and DSP targets on real-world office
environment human variability data (\cref{ssec:h_var}). Next, we evaluate the
QAT models on outdoor datasets (\cref{ssec:outdoor_var}) and scene-variability
(\cref{ssec:s_var}) in \cref{sec:performanceoutdoors} and
\cref{sec:performanescenevar} respectively.

\par The supplementary material provides a detailed comparison against
state-of-the-art event-based and RGB gesture-recognition methods, further
ablations of key model components and QAT training schemes, and an analysis of
performance generalisability beyond a single training instance.

\par Using ground truth labels, we quantified true-positive (TP), false-positive
(FP), and false-negative (FN) to calculate mean and median F1 scores across
microgesture classes, alongside power consumption and latency measurements for
both model configurations. For our experimental evaluation on real datasets, we
report results on three microgesture classes - swipe right (RS), swipe left (LS)
and combined pinch (CP, which is a combination of pinch and double-pinch
classes).
\subsection{Performance on Simulated Data} \label{sec:performancesimulateddata}
To evaluate our single-window (SW) low-power QAT model trained on simulated data, we
analysed its confusion matrix across all 10 gesture classes
(\Cref{fig:confusion_matrix_val}). The matrix's diagonal dominance shows
effective gesture discrimination, with minimal confusion between kinematically
similar gestures (e.g., 2.8\% between `swipe right (return)' and `left swipe').
The model achieved 87.9\% average class-wise precision, demonstrating effective
feature learning despite quantisation constraints. A t-SNE visualisation of the
learned feature embeddings is provided in the supplementary material.
\subsection{User Testing for Human Variability}
\label{sec:2channel_6channel_cpu_dsp}
\begin{table}[t]
  \centering
  \caption{Model performance on the Human Variability Study Dataset
  (\cref{ssec:h_var}) for Helios~1.0~\cite{bhattacharyya2024heliosextremelylowpower}
  and our single-window (2-ch) and temporal (6-ch) Helios~2.0 models. CPU models
  use the $4\times$ dataset, fine-tuned and QAT models the $4\times$ Augmented set.
  RS, LS, CP denote right swipe, left swipe, and combined pinch. Per-gesture
  median F1 is reported in the supplementary material.}
  \resizebox{\textwidth}{!}{%
    {%
      \small
      \renewcommand{\arraystretch}{1.1}
      \setlength{\tabcolsep}{4pt}
      \begin{tabular}{l|l|ccc|c|c}
        \hline
        \multirow{2}{*}{\textbf{Model}}
          & \multirow{2}{*}{\textbf{Type}}
          & \multicolumn{3}{c|}{\textbf{Mean F1}}
          & \multirow{2}{*}{\textbf{Power (mW)}}
          & \multirow{2}{*}{\textbf{Latency (ms)}} \\
        \cline{3-5}
          & & \textbf{RS} & \textbf{LS} & \textbf{CP} & & \\
        \hline
        Helios 1.0
          & CPU Model & 0.63 & 0.73 & 0.58 & 340 & 60   \\
        \hline
        \multirow{3}{*}{Single-window (SW)}
          & CPU Model      & 0.74 & 0.52 & 0.76 & 144 & 6.15 \\
          & Fine‑tuned CPU & 0.81 & 0.84 & 0.86 & 144 & 6.15 \\
          & QAT Model      & 0.82 & 0.82 & 0.84 &   6 & 2.35 \\
        \hline
        \multirow{3}{*}{Temporal (T)}
          & CPU Model      & 0.78 & 0.55 & 0.77 & 172 & 7.46 \\
          & Fine‑tuned CPU & 0.85 & 0.84 & 0.84 & 172 & 7.46 \\
          & QAT Model      & 0.81 & 0.84 & 0.89 &   8 & 4.60 \\
        \hline
      \end{tabular}
    }
  }
  \label{tab:model_channels_comparison}
  \vspace{-0.3cm}
\end{table}

\begin{table}[t]
  \centering
  \caption{Comparison of F1 Scores Across the single-window (SW) and temporal (T) QAT models and 3 Real Datasets: Human Variability (\cref{ssec:h_var}), Outdoors (\cref{ssec:outdoor_var} and Scene Variability (\cref{ssec:s_var}). All models are trained for 10 Epochs. Here RS stands for right swipe, LS is left swipe and CP is combined pinch.}
  \resizebox{\textwidth}{!}{%
    {%
      \small
      \renewcommand{\arraystretch}{1.2}
      \setlength{\tabcolsep}{4pt}
      \begin{tabular}{l|p{3cm}|*{5}{c|}c}
        \hline
        \multirow{2}{*}{\textbf{Model}}
          & \multirow{2}{=}{\textbf{Real Dataset}}
          & \multicolumn{3}{c|}{\textbf{Mean F1}}
          & \multicolumn{3}{c}{\textbf{Median F1}} \\
        \cline{3-8}
          & 
          & \textbf{RS} & \textbf{LS} & \textbf{CP}
          & \textbf{RS} & \textbf{LS} & \textbf{CP} \\
        \hline
        SW
          & Human Variability (Group~2)
          & 0.7365 & 0.7517 & 0.5439
          & 0.8856 & 0.8558 & 0.4259 \\
        \hline
        T
          & Human Variability (Group~2)
          & 0.7806 & 0.7540 & 0.7100
          & 0.9180 & 0.8576 & 0.6750 \\
        \hline
        SW
          & Outdoors (Group~2)
          & 0.8815 & \textbf{0.8440} & 0.7084
          & 0.8980 & \textbf{0.9189} & 0.8980 \\
        \hline
        T
          & Outdoors (Group~2)
          & \textbf{0.9468} & 0.8309 & \textbf{0.8375}
          & \textbf{0.9499} & 0.8743 & \textbf{0.9744} \\
        \hline\hline
        SW
          & Human Variability (User~3)
          & \textbf{0.9744} & \textbf{0.9756} & \textbf{1.0000}
          & \textbf{0.9744} & \textbf{0.9756} & \textbf{1.0000} \\
        \hline
        T
          & Human Variability (User~3)
          & \textbf{0.9744} & \textbf{0.9756} & \textbf{1.0000}
          & \textbf{0.9744} & \textbf{0.9756} & \textbf{1.0000} \\
        \hline
        SW
          & Scene Variability (User~3)
          & 0.8702 & 0.7496 & 0.6962
          & 0.8889 & 0.7242 & 0.7097 \\
        \hline
        T
          & Scene Variability (User~3)
          & 0.8789 & 0.8436 & 0.8363
          & 0.9041 & 0.8998 & 0.9189 \\
        \hline
      \end{tabular}
    }
  }
  \label{tab:allrealdataset}
  \vspace{-0.2cm}
\end{table}

In this section, we present results on real-world test data collected as
described in \cref{ssec:h_var} to evaluate our model's performance to human
variability.
\begin{figure}[t]
    \begin{minipage}[t]{0.48\linewidth}
	\centering
	\includegraphics[width=\linewidth]{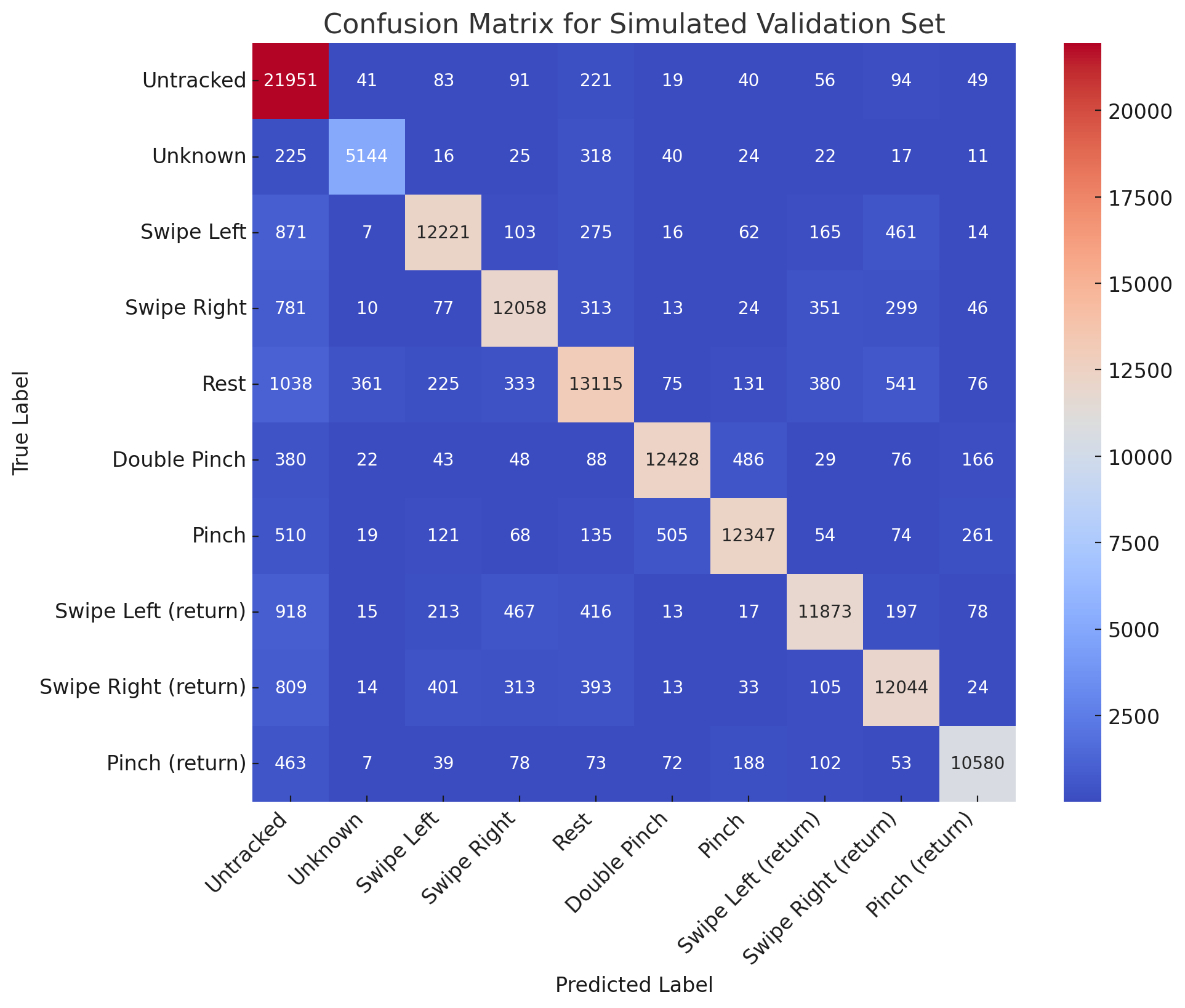}
	\vspace{-0.6cm}
	\caption{Confusion matrix for the single-window (SW) QAT model trained from
	scratch on simulated data. We report results on the simulated validation set
	across the 10 proposed classes.}
	\label{fig:confusion_matrix_val}
    \end{minipage}
    \hfill 
    \begin{minipage}[t]{0.48\linewidth}
	\centering
	\raisebox{0.5cm}{\includegraphics[width=\linewidth]{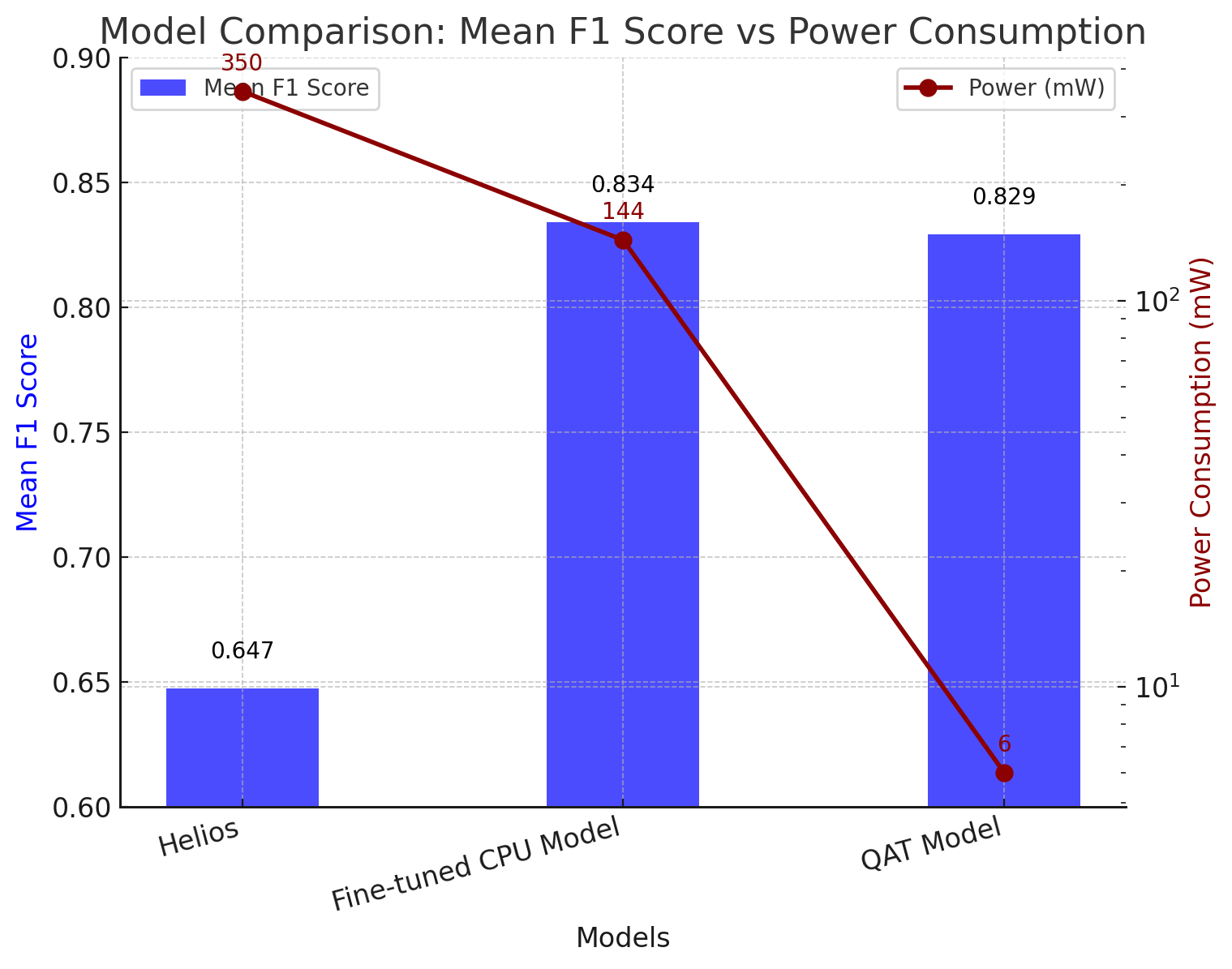}}
	\vspace{-0.6cm}
	\caption{Mean F1 score (across RS, LS and CP) for user averages plotted
	against power consumption for Helios 1.0
	\cite{bhattacharyya2024heliosextremelylowpower}, our finetuned CPU
	model, and our QAT model.}
	\label{fig:f1_scores_vs_power}
    \end{minipage}
\end{figure}
In \Cref{fig:f1_scores_vs_power}, we compare our Helios 2.0 CPU and QAT models
against the Helios 1.0 model presented in
\cite{bhattacharyya2024heliosextremelylowpower}. Given Helios 1.0 is a CPU only
model, comparing it our Helios 2.0 CPU model provides the most like-for-like
comparison. This shows that our model improves the F1 score from 0.647 to 0.834
as well as reducing the power requirement from 340mW to 144mW. It should be
noted however that the CPU and memory architecture are different.
\par
As we designed our Helios 2.0 model such that two of the five stages (two and
four) can be run on DSP we also compare to the QAT model. By quantising those
stages the F1 score reduces from 0.834 to 0.829, which is a minor reduction in
performance, whilst reducing the power to 6mW. This means our model improves F1
score by 30\% while only requiring 2.3\% of the power.
\subsection{User Testing for Environmental Variability}
\label{sec:performanceoutdoors} Comparing the Human Variability
(\cref{ssec:h_var}) and Outdoor (\cref{ssec:outdoor_var}) datasets in
\Cref{tab:allrealdataset} reveals consistent improvements outdoors. The
single-window (SW) model shows notable gains across all gestures, with Right
Swipe improving from 0.7365 to 0.8815 (mean F1) and Combined-Pinch from 0.4259
to 0.898 (median F1). Similarly, the temporal (T) model improves, particularly
for Right Swipe (0.7806 to 0.9468 mean F1) and Combined-Pinch (0.71 to 0.8375
mean F1, median F1 rising from 0.675 to 0.9744). Our quantised models thus
maintain, and often exceed, their controlled-environment performance outdoors,
likely due to stronger gesture signals against natural backgrounds and reduced
noise in brighter conditions.
\subsection{User Testing for Scene Variability} \label{sec:performanescenevar}
\Cref{tab:allrealdataset} shows a stark contrast between the simple office
environment of the Human Variability Dataset (\cref{ssec:h_var}) and the more
complex Scene Variability Dataset (\cref{ssec:s_var}). Both models achieve
near-perfect office scores (F1 of 0.97 for Right/Left Swipe and 1.0 for
Combined-Pinch), but scene variability degrades performance, especially for the
single-window (SW) model, whose Combined-Pinch drops to 0.6962 (mean F1). The
temporal (T) model is far more robust to changing backgrounds, with clear gains
for Left Swipe (0.8436 vs. 0.7496 mean F1) and Combined-Pinch (0.8363 vs. 0.6962
mean F1). This shows that background complexity in the training data matters, and
that the temporal model's extra context helps in difficult conditions.

\section{Conclusion}
\label{sec:conclusion}
In this paper, we present Helios 2.0, an ultra-low-power event-based vision
system for hand gesture control in smart glasses, addressing power efficiency,
adaptability, and user experience. Using a small set of microgestures that match
natural hand movements, an enhanced simulation methodology, a power-optimised
model architecture, and comprehensive benchmarking, we achieve state-of-the-art
F1 accuracy while operating at just 6-8~mW on a Qualcomm Snapdragon Hexagon DSP,
improving F1 by 20\% and reducing power by 25x over prior work. To our knowledge,
this is the first system to demonstrate accurate, low-power, low-latency gesture
recognition from event vision for next-generation smart glasses. In future work,
we aim to expand the gesture vocabulary, improve adaptation to individual users,
and integrate multi-modal sensing for practical, touch-free interaction on
wearable devices.

\clearpage
\appendix
\begin{center}\Large\bfseries Supplementary Material\end{center}
\vspace{0.3cm}

\section{Related Work} \label{sec:related}
While our work is the first to deliver an ultra-low-power solution for natural
hand interactions on smart glasses, it builds upon prior research in gesture
recognition, machine learning for event-based vision, and low-bit quantisation
for efficient inference. Below, we provide an overview of the most relevant
contributions in these areas.
\subsection{Hand Gesture Recognition} Hand Gesture Recognition (HGR) has been
explored across various sensing modalities~\cite{HGR-A-Review, HMISensing}.
Data gloves \cite{Dataglove} provide high accuracy but are unsuitable for
continuous use. Vision-based methods using traditional cameras
\cite{HandGRwLeap, StereoHGR3D, Liu2019KinectbasedHG, Bao2017TinyHG,
Singla2018VisualRO} combined with deep learning models \cite{JiIeeeTO,
Koller2016DeepHH, Molchanov2015HandGR, Molchanov2016OnlineDA,
Neverova2014MultiscaleDL, Sinha2016DeepHandRH} demonstrate strong performance in
controlled environments but struggle with environmental variations and
computational constraints. \cite{Chandra2016LowPG} reduced frame rates and
employed standby modes to save power to achieve real-time gesture recognition.
However, they faced a fundamental tradeoff: lower latency requires higher
sampling rates leading to increased power consumption. Alternative approaches
like sEMG \cite{semg} and ultrasound \cite{ultrasound} face challenges with
electrode placement, muscle fatigue, or specialised hardware requirements.  \par
\textbf{Our real-time gesture recognition contribution:} Compared to the
previous state-of-the-art, our approach improves power efficiency by 25$\times$
and gesture recognition accuracy by 20\%. This significantly advances the
feasibility of power-efficient real-time gesture recognition for deployment in
smart glasses.
\subsection{Event Camera Datasets: Real and Synthetic} Event cameras operate on
an asynchronous, per-pixel sensing mechanism, fundamentally different from
conventional frame-based cameras \cite{EventsSurveyPaper}. It is inspired by
the human retina where each pixel functions independently, continuously
detecting changes in light intensity. 
\par Real-world datasets,
captured using event cameras such as DAVIS \cite{DAVIS} and
\href{https://docs.prophesee.ai/stable/hw/evk/evk4.html}{Prophesee} cover
diverse applications like object detection, action recognition, tracking, and 3D
perception. EventVOT \cite{EventVOT} and eTraM \cite{etram} focus on object
tracking and traffic monitoring, while DVS-Lip \cite{DVS-Lip} and SeAct
\cite{SeAct} explore fine-grained human motion and action recognition.
Automotive datasets such as DSEC \cite{DSEC}, GEN1 \cite{Gen1}, 1 MPX
\cite{Prophesee_paper_detection_AVs} and N-Cars \cite{HATS+NCars} provide
event streams for autonomous driving. However, real-world datasets are
constrained by hardware limitations, environmental conditions, and annotation
challenges, making it difficult to scale data collection efficiently. To address
these gaps, synthetic datasets leverage simulators to generate large-scale,
controlled event streams. For example, Event-KITTI \cite{EventKITTI} extends
the popular KITTI dataset with event-based representations. Datasets such as
N-ImageNet \cite{NImagenet}, CIFAR10-DVS \cite{CIFAR10DVSAE}, and N-MNIST
\cite{NMNIST} convert standard image datasets into event streams, enabling
model benchmarking across classification tasks. Despite these advances,
synthetic datasets often lack realistic noise characteristics, which can impact
generalisation to real-world scenarios.  \par \textbf{Our datasets
contribution:} We propose the development of a specialised event-based simulator
to generate microgesture data tailored to our specific application. This
approach is motivated by the absence of both real-world and synthetic datasets
that sufficiently address our requirements. By leveraging simulation, we can
precisely control gesture variations, environmental conditions, and sensor
parameters, enabling the creation of high-fidelity data that aligns with our use
case. We also detail three datasets collected from internal studies, covering
user variability, where environment was fixed, environment variability, with a
single expert user across ambient lit complex scenes and an outdoor dataset
utilising one of the groups from the user variability study. These provide
benchmarks for key areas needed to access a system targeting smart-eyewear.
\subsection{Event Camera Simulators} A review of recent event simulators can be
found in \cite{recenteventcamerainnovations}. Notable simulators include the
DAVIS Simulator \cite{DAVISSimulator}, which generated event streams, intensity
frames, and depth maps with high temporal precision through time interpolation.
ESIM \cite{Rebecq18corl} extends this by modelling camera motion in 3D
environments. Unlike traditional frame-based simulators, ESIM accurately
simulates the asynchronous nature of event cameras, ensuring that events are
generated only when intensity changes occur. More advanced simulators like ICNS
Simulator \cite{ICNSSimulator} introduce realistic noise models improving the
fidelity of synthetic events, while DVS-Voltmeter \cite{DVS-Voltmeter}
incorporates stochastic variations in sensor behaviour.  \par \par \textbf{Our
simulation contribution:} Simulators like ESIM are not inherently designed for
human-centric applications, such as gesture recognition. To address this
limitation, we leverage the work of ESIM alongside our custom rendering engine,
developed in Unity, to generate synthetic event data of hands performing
microgestures.
\subsection{Event-Based Vision} ML approaches for event-based classification and
detection typically transform raw event data into structured representations,
such as time surfaces or event volumes, which can then be processed using
standard Convolutional Neural Networks (CNNs) or recurrent architectures like
Conv-LSTMs \cite{millerdurai2024eventego3d, liang2024towards, gao2024sd2event,
chen2024segment, kong2024openess, aliminati2024sevd, EventstoVideo_19}. While
these methods are straightforward to implement, they introduce redundant
computations and can be computationally expensive due to the dense processing of
inherently sparse data. To overcome these inefficiencies, researchers have
explored sparse convolutional architectures \cite{liu2015sparse} which compute
convolutions only at active sites with non-zero feature vectors, significantly
reducing computational overhead \cite{peng2024scene, yu2024eventps,
zhang2024co}. Additionally, techniques leveraging temporal sparsity aim to
retain previous activations and apply recursive sparse updates
\cite{ren2024simple, SparseEventGraphGNN}, rather than reprocessing the entire
event volume. Beyond grid-based representations, graph-based methods maintain
the compact, asynchronous structure of event streams, with Graph Neural Networks
(GNNs) providing a framework for handling irregular and dynamic event data
\cite{SparseEventGraphGNN, GNNsforEvents}. However, this method still takes
202ms when implemented in Python and CUDA and run on Nvidia Quadro RTX. Current
event-based ML methods are therefore still severely limited for on-device
deployment.  
traditional synchronous, dense images \cite{EventsSurveyPaper}, necessitating
specialised representations and architectures.  Quantisation techniques reduce
neural network latency and power consumption by storing weights and activation
tensors in lower (8-bit) precision \cite{WhitePaperonNeuralNetworkQuantization,
IntegerQDeepLearningInference, QGaitTA}. Post-Training Quantisation (PTQ)
\cite{PostTrainingQO} quantises pre-trained models with minimal engineering
effort, while Quantisation-Aware Training (QAT) \cite{QuantizationAT}
incorporates simulated quantisation during training to better preserve accuracy,
especially for low-bit quantisation.  \par \textbf{Our machine learning
contribution:} We develop a QAT-scheme for our custom architecture to
significantly reduce power consumption without sacrificing gesture recognition
accuracy. Helios 2.0 reduces power consumption by $25 \times$ (from 150~mW DSP-equivalent
to 6~mW), lowers latency from 60~ms to 2.35~ms, and improves F1 accuracy by 20\%
over Helios~1.0~\cite{bhattacharyya2024heliosextremelylowpower}. Compared to
Helios 1.0, Helios 2.0 uses longer training sequences spanning 2~s with six
gestures per sequence (vs. one gesture per sequence in Helios 1.0), along with
structured Markov-based transitions, class-balanced sampling, and kinematic
motion blending for more realistic and diverse simulation data. Unlike Helios
1.0, Helios 2.0 introduces a five-stage, quantisation-aware architecture with
$>$99.8\% of compute optimised for low-power DSP execution. We also redesign 
the training approach to use longer sequences and learn gesture class 
probabilities through thresholded aggregation. Additionally, we apply
rotational augmentation during fine-tuning to improve robustness to hand pose
variations. We acknowledge \cite{GesturewithEvents_2017} as the closest related
benchmark, but note that it involves large, deliberate gestures, making it a
significantly easier and less realistic task for smart wearable interaction. In
contrast, our work focuses on short, subtle microgestures suitable for real-time
application control.
Our contribution fills a critical gap in benchmarking low-power, event-based
wearable interfaces. Unlike Meta's approach \cite{STMG}, which relies on
frame-based cameras and hand skeletal tracking, our method eliminates the need
for explicit hand pose estimation by directly recognising gestures from raw
event streams. \cite{GesturewithEvents_2017} also utilises event-based gesture
recognition but employs the DVS-128 camera, which is impractical for integration
with smart eyewear. Additionally, we address the critical challenge of false
positives from ego-motion by explicitly modelling adversarial classes alongside
relevant microgestures, a concern not actively mitigated in prior works
\cite{aliminati2024sevd, GesturewithEvents_2017, STMG}.
\subsection{Overview of Event Cameras} \label{subsec:overvieweventcameras}
Event sensors offer unprecedented temporal resolution while operating at
extremely low power levels, as low as 3~mW. In contrast, traditional frame-based
sensors require a trade-off between temporal resolution and power consumption,
typically ranging from 35 to 200~mW depending on the frame rate. Event cameras
use an asynchronous, per-pixel sensing mechanism, fundamentally different from
conventional frame-based cameras \cite{EventsSurveyPaper}, and are inspired by
the human retina where each pixel functions independently, continuously detecting
changes in light intensity. When the intensity at an event-camera pixel surpasses
a predefined threshold, the sensor records an event $\langle x,y,t,p \rangle$,
where $(x,y)$ are the pixel coordinates, $t$ is the event timestamp, and
$p \in \{1, -1\}$ indicates whether the light intensity increased or decreased.
While conventional vision architectures are optimised for frame-based data, they
struggle to meet the demands of wearable, low-power applications, motivating the
event-based methods surveyed above.
\subsection{Event Representations}
Two widely used representations of event data for machine learning applications
are time surfaces and event volumes \cite{EventsSurveyPaper}. Time surfaces store
the timestamp of the most recent event for each pixel and polarity, forming a 2D
map where each pixel retains a single time value. This compact representation
efficiently encodes temporal dynamics but becomes less effective in textured
scenes where pixels generate frequent events. By aggregating local memory time
surfaces, Histograms of Averaged Time Surfaces (HATS) \cite{HATS+NCars} construct
a higher-order representation to improve temporal and noise robustness. In
contrast, event volumes represent events as 3D histograms, maintaining richer
temporal structure by accumulating events over time; while more detailed, they
may lose polarity information due to voxel-based accumulation, reducing their
effectiveness in tasks requiring fine-grained contrast changes. In our work, we
employ polarity-separated time surfaces \cite{rudnev2021eventhands} for both
computational and storage efficiency: separating polarities mitigates event
clashes and yields a more structured, accurate representation of the temporal
dynamics in event data.

\section{Comparison with State-of-the-Art Methods} \label{sec:sota_comparison}
This section compares Helios 2.0 with existing state-of-the-art event-based and
RGB-based gesture recognition methods.

\subsection{Event-Based Vision Methods} We compare Helios 2.0 against recent
event-based vision approaches for gesture recognition and related tasks.
\Cref{tab:baseline_comparison} compares them across inference time, hardware
requirements, and power consumption.

\begin{table}[t]
\centering
\caption{Comparison of event-based vision methods for gesture recognition and related tasks.}
\label{tab:baseline_comparison}
\small
\setlength{\tabcolsep}{4pt}
\renewcommand{\arraystretch}{1.15}
\resizebox{\textwidth}{!}{%
\begin{tabular}{p{2.4cm}|c|p{1.7cm}|c|p{1.9cm}|c|p{1.6cm}}
\hline
\textbf{Method} & \textbf{Year} & \textbf{Arch.} & \textbf{Inf.\ time} & \textbf{Hardware} & \textbf{Power} & \textbf{Task} \\
\hline
\textbf{\cite{GesturewithEvents_2017}} & CVPR 2017 & CNN & 105ms & TrueNorth & <200mW & Large deliberate gestures* \\
Events-to-Video \cite{EventstoVideo_19} & CVPR 2019 & UNet + Recurrent & <10ms & RTX 2080 Ti & - & Object classification \\
Event-based Async Sparse Conv \cite{SparseEventConvs} & ECCV 2020 & Sparse CNN & 80.4ms & i7-6900K CPU & - & Object detection \\
Learning to Detect Objects \cite{Prophesee_paper_detection_AVs} & NeurIPS 2020 & ConvLSTM & 16.7-39.3ms & GTX 980 & - & Object detection \\
EventHands \cite{rudnev2021eventhands} & ICCV 2021 & ResNet-18 & 0.65-1.3ms$^\dagger$ & GTX 2070/RTX 2080 Ti & - & Hand pose (3D joints) \\
AEGNN \cite{SparseEventGraphGNN} & CVPR 2022 & Graph Neural Net & 92-167ms & Quadro RTX & - & Object detection \\
Efficient Human Pose \cite{EventsasPointCloud} & 3DV 2022 & PointNet/DGCNN & 12.29ms & Jetson Xavier NX & - & Human pose (2D) \\
RVT \cite{RVT} & CVPR 2023 & Transformer+LSTM & $\sim$10ms & T4 GPU & - & Object detection \\
Data-driven Feature Tracking \cite{IlluminationandEvents} & CVPR 2023 & Conv-LSTM+Attention & 17ms & Quadro RTX 8000 & - & Feature tracking \\
SAST \cite{peng2024scene} & CVPR 2024 & Sparse Transformer+LSTM & 14.5-19.7ms & TITAN Xp & - & Object detection \\
\hline
\textbf{Helios 1.0} \cite{bhattacharyya2024heliosextremelylowpower} & 2024 & CNN & 60ms & CPU & 340mW & Microgestures \\
\textbf{Helios 2.0 (CPU)} & 2024 & CNN & 6.15ms & Snapdragon XR2Gen2 & 144mW & Microgestures \\
\textbf{Helios 2.0 (QAT)} & 2024 & Quantized CNN & \textbf{2.35ms} & Snapdragon XR2Gen2 & \textbf{6mW} & Microgestures \\
\hline
\end{tabular}%
}

\vspace{0.2cm}
\footnotesize{*Large, deliberate gestures represent a significantly easier and less realistic task for smart wearable interaction compared to microgestures.}

\footnotesize{$^\dagger$EventHands reports 750-1550 poses/second throughput.}
\end{table}

Compared to prior work, Helios 2.0 improves accuracy by over 20\% while lowering power and latency. Specifically, we achieve:
\begin{itemize}
    \item 44$\times$ faster inference than \cite{GesturewithEvents_2017} (2.35ms vs 105ms)
    \item 25$\times$ lower power than Helios 1.0 (6mW vs 150mW DSP-equivalent)
    \item A focus on challenging microgestures rather than large, deliberate gestures
\end{itemize}

While event-based gesture recognition has been explored before, our combination of microgesture recognition and low power makes the system suitable for always-on wearable use.

\subsection{RGB-Based Vision Methods}
We also compare against RGB-based approaches to demonstrate the fundamental advantages of event-based vision for wearable applications:

\begin{itemize}
    \item \cite{molchanov2015hand}: Uses 3D CNNs on RGB and depth data for driver gesture recognition. Achieves 12.5ms inference on GPU (400 FPS for low-resolution network) and 78ms on CPU. Their high-resolution network requires 68ms on GPU, demonstrating the computational requirements of RGB processing.
    \item \cite{kopuklu2019real}: Implements real-time hand gesture detection using CNNs on RGB input. Achieves 62 FPS with ResNeXt-101 (41 FPS with C3D) when both detector and classifier are active on NVIDIA Titan Xp GPU, translating to approximately 16-24ms latency per frame.
\end{itemize}

\begin{table}[h]
\centering
\caption{Comparison with RGB-based gesture recognition methods}
\label{tab:rgb_comparison}
\begin{tabular}{l|c|c|c}
\hline
\textbf{Method} & \textbf{Latency} & \textbf{Hardware} & \textbf{Power} \\
\hline
\cite{molchanov2015hand} (RGB) & 12.5ms & GPU & >10W* \\
\cite{molchanov2015hand} (RGB) & 78ms & CPU (i5) & $\sim$15W* \\
\cite{kopuklu2019real} (RGB) & 16-24ms & Titan Xp GPU & $\sim$250W* \\
\cite{GesturewithEvents_2017} (Event) & 105ms & TrueNorth & <200mW \\
\textbf{Helios 2.0 (Event)} & \textbf{2.35ms} & \textbf{DSP} & \textbf{6mW} \\
\hline
\end{tabular}

\vspace{0.2cm}
\footnotesize{*Typical GPU/CPU power consumption}
\end{table}

While RGB methods can achieve low latency with powerful hardware, they require 3-4 orders of magnitude more power than our event-based approach. For wearable microgesture recognition, our approach provides both lower latency (2.35ms vs 12.5ms best-case RGB) and far lower power (6mW vs 10-250W), making it suitable for always-on wearable operation.
\FloatBarrier

\section{Additional Dataset Illustrations}
This section illustrates the Monk Skin Tone Scale used to characterise
participant skin tone across our benchmark datasets.

\begin{figure}[h]
    \centering
    \includegraphics[height=4cm]{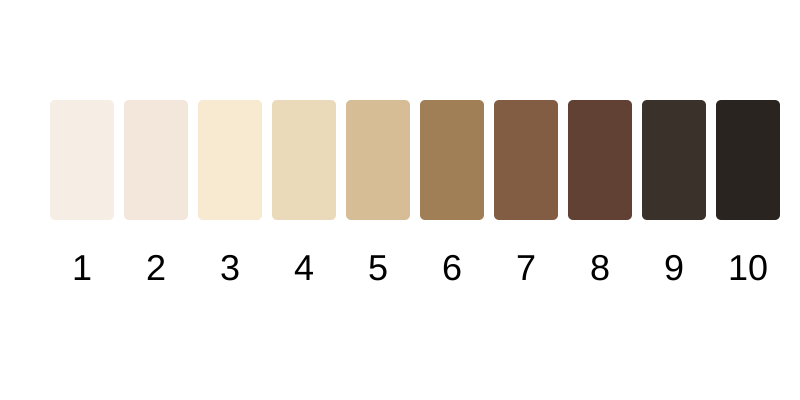}
    \caption{The Monk Skin Tone Scale, with corresponding numbering from 1 light
    to 10 dark.} \label{fig:msk}
\end{figure}

\section{Training Parametrisation and Inference Details} \label{trainingdetails}
We trained the network for 10 epochs with a batch size of 512. We use the Adam
\cite{adam} optimiser with an initial learning rate of 0.0005. We utilise a
learning rate scheduler, which linearly decays the learning rate after an
initial period where the learning rate is held fixed. We apply dropout with 0.2
drop rate on the dense layers and set the gesture threshold value to 0.6.  \par
At inference time the model takes as input TS of window size 240~ms. To avoid
sampling issues we follow the approach of Helios~1.0
\cite{bhattacharyya2024heliosextremelylowpower} and create a TS every 80~ms.
This follows the intuition that the model will be less confident on predictions
where non-complete microgestures have occurred in the input frame. The model
will therefore predict a microgesture with a larger probability when a complete
microgesture has occurred and a lower probability when a non-complete
microgesture has occurred. We also overcome the issue of the model predicting a
microgesture with low confidence by introducing a threshold on softmax
probabilities, at which a microgesture is deemed to have occurred, and set it to
0.65.

\subsection{Loss Function} Following Helios~1.0
\cite{bhattacharyya2024heliosextremelylowpower}, we employ two loss functions:
$\mathcal{L}_{\text{bbox}}$, which calculates mean squared error between
predicted and true bounding boxes (only when hands are present), and
$\mathcal{L}_{\text{gesture}}$, a sparse categorical cross entropy loss for
microgesture classification. $\mathcal{L}_{\text{bbox}}$ is essential because
$\mathcal{L}_{\text{gesture}}$ alone cannot effectively guide hand centring in
TS crops, and without it, the model initially focuses solely on distinguishing
the no-hand class before learning other microgestures, leading to suboptimal
performance due to local minima.

\section{Ablation Studies} \label{sec: ablation}
Our ablation study (Table~1 in the main paper) shows that the single-window (SW)
and temporal (T) models benefit from fine-tuning with rotation augmentation. This
improvement occurs because the pre-training dataset, while useful for learning
multiple micro-gesture sequences, lacks natural hand orientations. Fine-tuning
helps the model adapt to realistic human hand distributions. While the temporal
(T) model yields modest performance gains, it requires triple the data
throughput, potentially increasing system power demands.
\begin{table}[t]
  \centering
  \caption{Ablation of quantisation-aware training (QAT) model components,
  evaluated on the Human Variability Study Dataset (main paper). We ablate
  architectural choices (Conv2D, depthwise-separable Conv2D, Dense layers,
  Conv1D, Global Average Pooling (GAP), and strided downsampling) across the five
  stages of our model, using the single-window (SW) QAT model as the control.}
  \small
  \renewcommand{\arraystretch}{1.15}
  \setlength{\tabcolsep}{4pt}
  \resizebox{\textwidth}{!}{%
  \begin{tabular}{p{4.5cm}|c|ccc|ccc|c}
    \hline
    \textbf{Model} & \textbf{Params, FLOPs}
      & \multicolumn{3}{c|}{\textbf{Mean F1}}
      & \multicolumn{3}{c|}{\textbf{Median F1}}
      & \textbf{Power} \\
    \cline{3-5}\cline{6-8}
      & & RS & LS & CP & RS & LS & CP & \textbf{(mW)} \\
    \hline\hline
    Single-window (SW) (control)
      & 596k, 164M & 0.82 & \textbf{0.82} & 0.84 & 0.90 & \textbf{0.88} & 0.95 & 6 \\
    \hline
    Add 2 Dense layers in Stage 5
      & 604k, 164M & 0.80 & 0.80 & 0.86 & 0.91 & 0.86 & 0.94 & 5 \\
    \hline
    Conv1D in place of Dense in Stages 2\&4
      & 605k, 165M & \textbf{0.82} & 0.81 & \textbf{0.87} & \textbf{0.93} & 0.88 & 0.95 & 5 \\
    \hline
    Depthwise Separable Conv2D in Stages 2\&4
      & 1.2M, 55M & 0.77 & 0.80 & 0.79 & 0.90 & 0.88 & \textbf{0.97} & 6 \\
    \hline
    Depthwise Separable Conv2D + strided downsampling in Stages 2\&4
      & 391k, 25M & 0.76 & 0.62 & 0.74 & 0.85 & 0.66 & 0.92 & \textbf{4} \\
    \hline
    Depthwise Separable Conv2D + Conv1D (Dense$\rightarrow$Conv1D) in Stages 2\&4
      & \textbf{113k, 24M} & 0.75 & 0.78 & 0.75 & 0.85 & 0.86 & 0.90 & \textbf{4} \\
    \hline
    Depthwise Separable Conv2D + GAP (Dense$\rightarrow$GAP) in Stages 2\&4
      & 70k, 52M & 0.41 & 0.23 & 0.55 & 0.29 & 0.18 & 0.66 & \textbf{4} \\
    \hline
    Depthwise Separable Conv2D + Conv1D (stride=4) in Stages 2\&4
      & 86k, 53M & 0.64 & 0.70 & 0.74 & 0.67 & 0.81 & 0.89 & \textbf{4} \\
    \hline
  \end{tabular}%
  }
  \label{tab:qat_components_ablate}
  \vspace{-0.2cm}
\end{table}

\par
\Cref{tab:qat_components_ablate} compares various model modifications. Adding
extra dense layers or replacing them with conv1D produced mixed results with
minimal impact, suggesting small model capacity increases and layer type have
limited effect on performance. Replacing conv2D with depthwise separable conv2D
reduced FLOPs but showed inconsistent results, with decreasing mean F1 scores
while slightly improving median scores. This indicates better performance for
already well-served users at the expense of others. Other configurations
(strided downsampling, depthwise separable conv2D with conv1D, global average
pooling, and larger conv1D strides) all reduced both model size and performance,
indicating a minimum capacity threshold required for effective micro-gesture
learning.
\FloatBarrier

\section{QAT Training Strategies} \label{sec:QAT training strategy}
\begin{table}[t]
    \centering
    \caption{Impact of different training strategies for quantisation-aware
    training (QAT), evaluated on the Human Variability Study Dataset (main paper).
    Strategies include training QAT from scratch; training a base model then QAT;
    and training a base model, fine-tuning, then QAT. RS is right swipe, LS is
    left swipe, and CP is combined pinch.}
    \small
    \renewcommand{\arraystretch}{1.2}
    \setlength{\tabcolsep}{4pt}
    \resizebox{\textwidth}{!}{%
    \begin{tabular}{p{2.7cm}|l|c|c|c|c|c|c|c|c}
        \hline
        \textbf{QAT Training Strategy} & \textbf{Stage} & \textbf{Epochs} & \textbf{Dataset} & \textbf{F1$_\text{RS}$} & \textbf{F1$_\text{LS}$} & \textbf{F1$_\text{CP}$} & \textbf{Med$_\text{RS}$} & \textbf{Med$_\text{LS}$} & \textbf{Med$_\text{CP}$} \\
        \hline
        \hline
        From scratch & QAT & 10 & $4\times$ Aug. & \textbf{0.82} & 0.82 & 0.84 & 0.90 & 0.88 & 0.95 \\
        \hline
        \multirow{2}{2.7cm}{From $1\times$ Base Model} & Base & 10 & $1\times$ & \multirow{2}{*}{0.78} & \multirow{2}{*}{0.81} & \multirow{2}{*}{\textbf{0.86}} & \multirow{2}{*}{0.92} & \multirow{2}{*}{0.89} & \multirow{2}{*}{0.96} \\
        & QAT & 5 & $4\times$ Aug. &  &  &  &  &  &  \\
        \hline
        \multirow{3}{2.7cm}{From $4\times$ Fine-tuned $1\times$ Base Model} & Base & 10 & $1\times$ & \multirow{3}{*}{0.77} & \multirow{3}{*}{\textbf{0.83}} & \multirow{3}{*}{0.85} & \multirow{3}{*}{0.91} & \multirow{3}{*}{\textbf{0.90}} & \multirow{3}{*}{0.97} \\
        & Fine-tuned & 5 & $4\times$ Aug. &  &  &  &  &  &  \\
        & QAT & 5 & $4\times$ Aug. &  &  &  &  &  &  \\
        \hline
        \multirow{3}{2.7cm}{From $4\times$ Fine-tuned $4\times$ Base Model} & Base & 10 & $4\times$ & \multirow{3}{*}{0.82} & \multirow{3}{*}{0.80} & \multirow{3}{*}{0.86} & \multirow{3}{*}{\textbf{0.94}} & \multirow{3}{*}{0.88} & \multirow{3}{*}{\textbf{1.00}} \\
        & Fine-tuned & 5 & $4\times$ Aug. &  &  &  &  &  &  \\
        & QAT & 5 & $4\times$ Aug. &  &  &  &  &  &  \\
        \hline
    \end{tabular}%
    }
    \label{tab:model_comparison}
\end{table}

QAT can be set up in different ways: (a) training a model from scratch;
(b) training a base model and then applying QAT; and (c) training a base model,
fine-tuning on augmented data, and then applying QAT. \Cref{tab:model_comparison}
shows the impact of these strategies. Based on the mean F1 scores across all
three microgestures, training from scratch and training a $4\times$ base model
followed by $4\times$ fine-tuning perform the strongest, both scoring above 0.8
F1 across all microgestures. When considering the median scores, the model
trained from a $4\times$ base model and then fine-tuned on $4\times$ data
achieves the highest average F1 across the three microgestures.
\FloatBarrier

\section{QAT Model Performance and Seed Variability} \label{sec:QAT training
variability}
\begin{table}[t]
  \centering
  \caption{Median F1 on the Human Variability Study Dataset for
  Helios~1.0~\cite{bhattacharyya2024heliosextremelylowpower} and our single-window
  (SW) and temporal (T) Helios~2.0 models, complementing the mean-F1 values in the
  main paper Table~1. CPU models use the $4\times$ dataset, fine-tuned and QAT
  models the $4\times$ Augmented set. RS, LS, CP denote right swipe, left swipe,
  and combined pinch.}
  \resizebox{\textwidth}{!}{%
    {%
      \small
      \renewcommand{\arraystretch}{1.1}
      \setlength{\tabcolsep}{4pt}
      \begin{tabular}{l|l|ccc|c|c}
        \hline
        \multirow{2}{*}{\textbf{Model}} & \multirow{2}{*}{\textbf{Type}}
          & \multicolumn{3}{c|}{\textbf{Median F1}}
          & \multirow{2}{*}{\textbf{Power (mW)}} & \multirow{2}{*}{\textbf{Latency (ms)}} \\
        \cline{3-5}
          & & \textbf{RS} & \textbf{LS} & \textbf{CP} & & \\
        \hline
        Helios 1.0 & CPU Model & 0.68 & 0.87 & 0.60 & 340 & 60 \\
        \hline
        \multirow{3}{*}{Single-window (SW)}
          & CPU Model      & 0.72 & 0.57 & 0.86 & 144 & 6.15 \\
          & Fine-tuned CPU & 0.90 & 0.91 & 0.95 & 144 & 6.15 \\
          & QAT Model      & 0.90 & 0.88 & 0.95 &   6 & 2.35 \\
        \hline
        \multirow{3}{*}{Temporal (T)}
          & CPU Model      & 0.76 & 0.66 & 0.87 & 172 & 7.46 \\
          & Fine-tuned CPU & 0.93 & 0.91 & 0.92 & 172 & 7.46 \\
          & QAT Model      & 0.93 & 0.89 & 0.95 &   8 & 4.60 \\
        \hline
      \end{tabular}
    }
  }
  \label{tab:human_var_median}
  \vspace{-0.3cm}
\end{table}

\begin{figure}[ht]
    \centering
    \includegraphics[height=5cm]{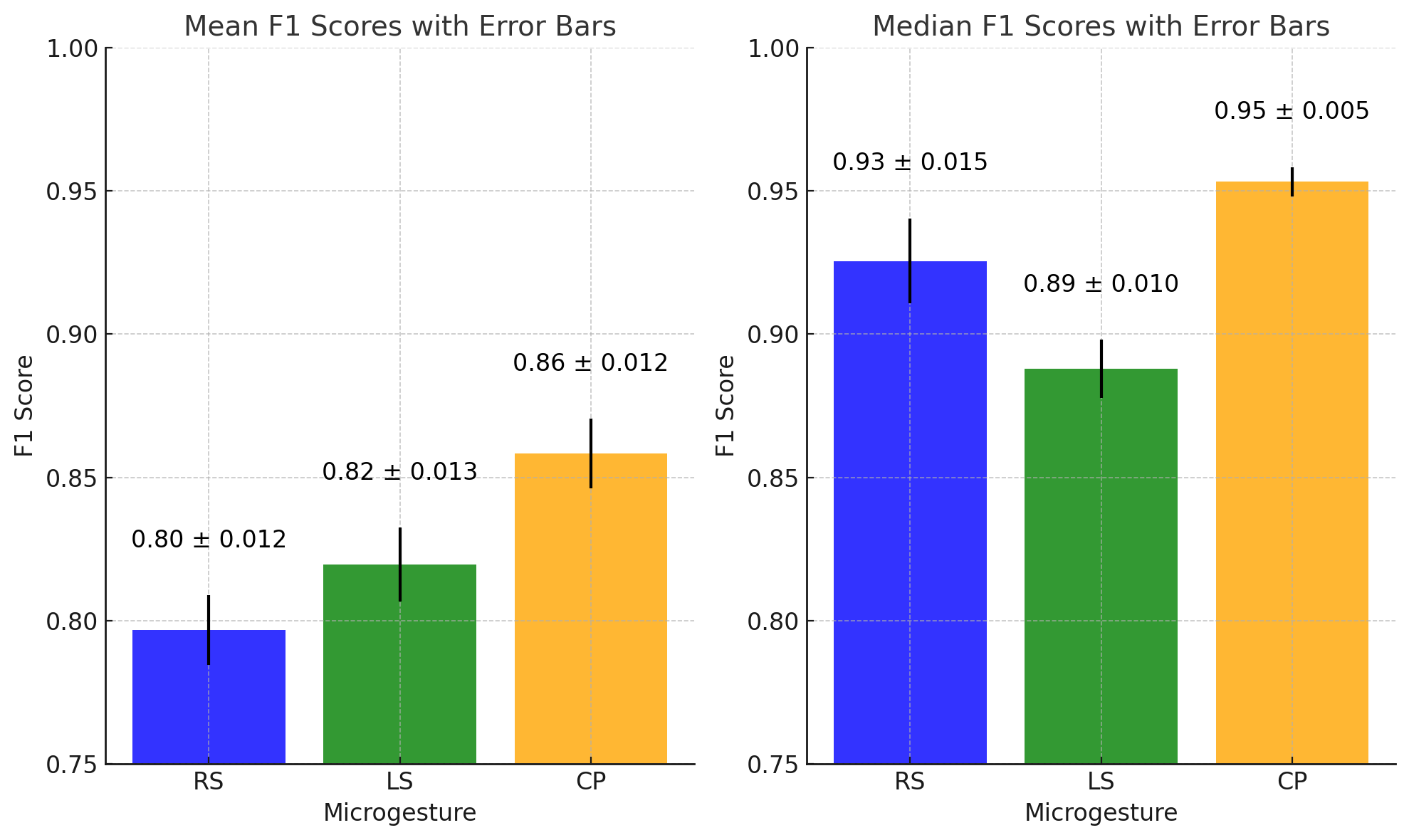}
    \caption{Mean and median F1 scores with error bars across 5 seeds for
    microgestures, training single-window (SW) QAT models.}
    \label{fig:f1_scores_sweep} \vspace{-0.2cm}
\end{figure}

\Cref{tab:human_var_median} reports the per-gesture median F1 for the Human
Variability Study, complementing the mean F1 in the main paper Table~1.
We analyse seed variability using standard deviation $\sigma$ as an error measure
in \Cref{fig:f1_scores_sweep}. We compute $\sigma$ across five independent runs
for both mean and median F1 scores. Seed variability for QAT models is important
for assessing generalisability beyond a single training instance. We observe that
both mean and median F1 scores have small error bars, suggesting that the model
performs consistently across runs. CP exhibits the lowest median variability
($\sigma = 0.005$), making it the most stable gesture. LS follows with slightly
higher median variability ($\sigma = 0.010$), while RS shows the highest
fluctuation ($\sigma = 0.015$). This contrast highlights a difference in
recognition stability: CP remains stable throughout the tests, while RS is more
sensitive to seed initialisation.
\FloatBarrier
\section{Helios 2.0 t-SNE Visualisation of Learned Feature Embeddings}
\begin{figure}[t]
    \centering
    \includegraphics[height=10cm]{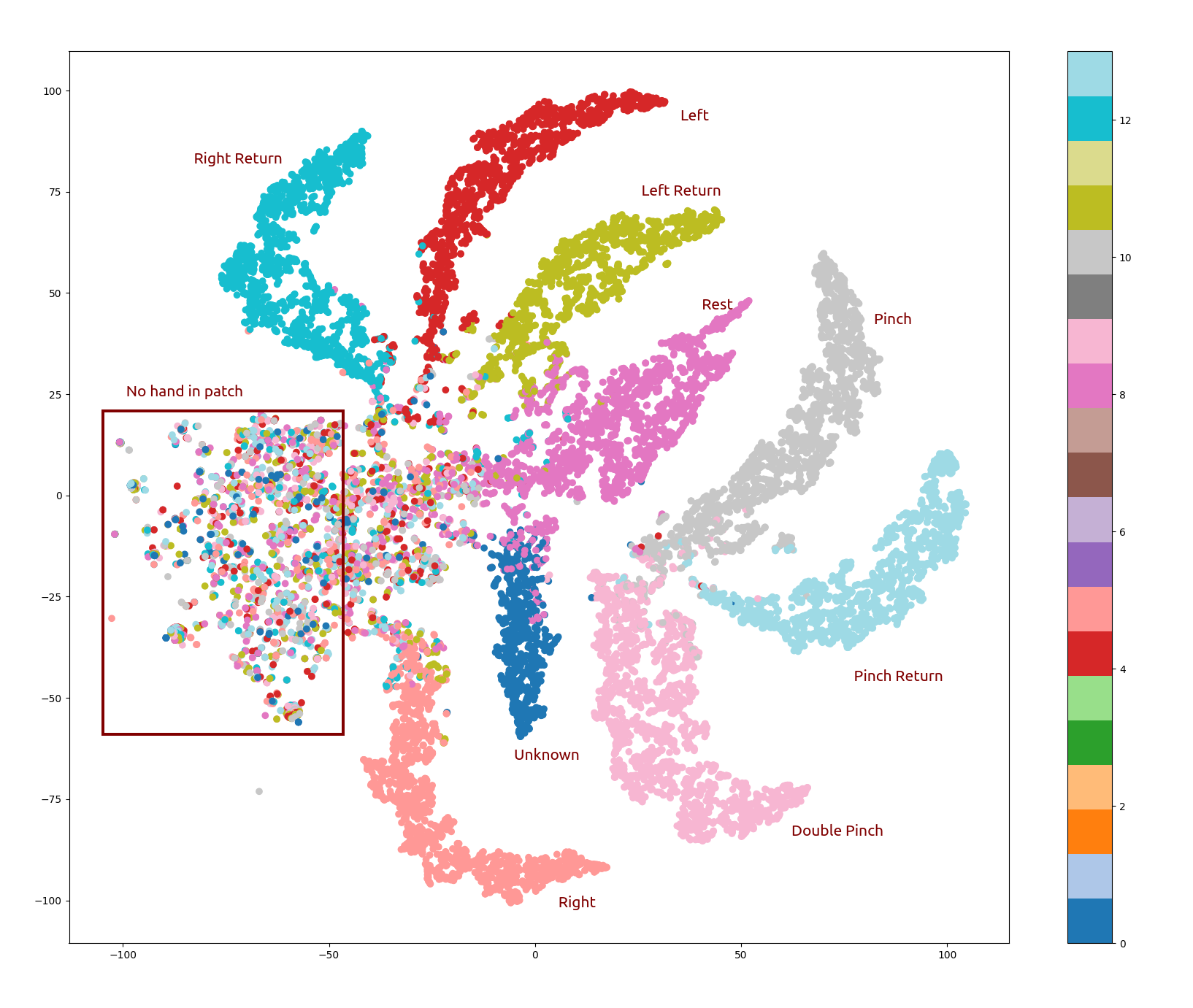}
    \caption{t-SNE for the single-window (SW) fine-tuned CPU model across 10
    classes.}
    \label{fig:tsne}
\end{figure}

\Cref{fig:tsne} illustrates the t-SNE visualisation of our single-window (SW) QAT
model's learned feature embeddings from the dense layer in Stage 5, across the 10
microgesture classes. The visualisation reveals distinct clustering patterns for
each gesture type, demonstrating the model's ability to learn discriminative
representations despite the quantisation constraints. Directional gestures such as
`Left', `Right', `Left Return', and `Right Return' form well-separated clusters
with clear trajectory patterns, while `Pinch', `Double Pinch', and `Pinch Return'
show structured separation with minimal overlap. The `Rest' class manifests as a
cluster with moderate dispersion. The `No hand in patch' region (highlighted by
the red box) shows a mixed distribution of points, indicating the model's
uncertainty when no clear hand features are present, and the `Unknown' class forms
a concentrated cluster, suggesting consistent identification of ambiguous hand
motions. This embedding structure provides visual evidence of the model's feature
extraction, which contributes to the classification performance observed in our
quantitative evaluations.

\section{Ethics Statement}
All participants in the human variability, scene variability, and outdoor
performance studies were employees of the authors' organisation who voluntarily
participated after being informed of the study's purpose and its intended use in
research publication. Participants were made aware of the data being collected
(hand size measurements, skin tone classification using the Monk Skin Tone Scale,
and raw event stream recordings) and provided verbal consent prior to
participation. Participant data, including names, was stored on the
organisation's internal systems to enable follow-up testing with updated model
iterations. All participant data is anonymized in this publication and referred
to only by anonymous identifiers. No personally identifiable images of
participants were captured; all images of hands appearing in this paper are of a
consenting author. Participants were free to withdraw at any time without
consequence. The planning, conduct, and reporting of this research are consistent
with widely recognised ethical principles for research involving human
participants.

\nocite{Berend2020,CIFAR10DVSAE,Chandra2016LowPG,DAVIS,DAVISSimulator,DSEC,DVS-Lip,DVS-Voltmeter,Dataglove,EventKITTI,EventVOT,EventsSurveyPaper,EventsasPointCloud,EventstoVideo_19,GNNsforEvents,Gen1,GesturewithEvents_2017,HATS+NCars,HGR-A-Review,HMISensing,ICNSSimulator,IlluminationandEvents,NImagenet,NMNIST,PostTrainingQO,Prophesee_paper_detection_AVs,QuantizationAT,RVT,Rebecq18corl,STMG,SeAct,SparseEventConvs,SparseEventGraphGNN,adam,aliminati2024sevd,bhattacharyya2024heliosextremelylowpower,etram,kopuklu2019real,liu2015sparse,molchanov2015hand,mst,peng2024scene,qualcommSoCpage,recenteventcamerainnovations,ren2024simple,rudnev2021eventhands,semg,ul-orion,ultrasound,unity}

\bibliographystyle{splncs04}
\bibliography{sample-base}

\end{document}